\documentclass[MSNbibl,nameyear,seceqn,dvips]{arxstspdf}
\usepackage{flushend}
\usepackage{stfloats}
\usepackage{graphicx}
\volume{29}
\issue{2}
\pubyear{2014}
\firstpage{302}
\lastpage{321}
\doi{10.1214/13-STS456} %kopijuoti is PTS


\begin{document}
\begin{frontmatter}

\title{Pooled Association Tests for Rare Genetic Variants: A Review and
Some New Results}%\thanksref{T1}
% kai straipsnis turi susijusiu diskusiju ir rejoinder'iu
%rejoinder at \relateddoi{r}{10.1214/00-STSXXXX}.}
\runtitle{Unified Framework for Analyzing Rare Variants}

\begin{aug}
\author[a]{\fnms{Andriy} \snm{Derkach}},
\author[b]{\fnms{Jerry F.} \snm{Lawless}}
\and
\author[c]{\fnms{Lei} \snm{Sun}\corref{}\ead[label=e1]{sun@utstat.toronto.edu}}
\runauthor{A. Derkach, J.~F. Lawless and L. Sun}
%University of Waterloo, Waterloo ON N2L 3G1.}
%Health, University of Toronto, 155 College Street, Toronto, ON M5T 3M7.}
\affiliation{University of Toronto, University of Waterloo and
University of Toronto,
and University of Toronto}

\address[a]{Andriy Derkach is Graduate Student, Department of Statistical
Sciences,
University of Toronto,
100 St. George Street, Toronto, Ontario, Canada 1M5S 3G3.}
\address[b]{Jerry F. Lawless is Distinguished
Professor
Emeritus, Department of Statistics
and Actuarial Science,
University of Waterloo, Waterloo, Ontario, Canada N2L 3G1 and Professor,
Division of Biostatistics, Dalla Lana School
of Public Health, University of Toronto, 155 College Street, Toronto,
Ontario, Canada M5T 3M7.}
\address[c]{Lei Sun is Associate
Professor, Department of Statistical Sciences,
University of Toronto,
100 St. George Street, Toronto, Ontario, Canada 1M5S 3G3 and Associate
Professor,
Division of Biostatistics, Dalla Lana School of Public Health,
University of Toronto, 155 College Street, Toronto, Ontario, Canada M5T 3M7 \printead{e1}.}
\end{aug}

% ABSTRACT
%
\begin{abstract}
In the search for genetic factors that are associated with complex
heritable human traits, considerable attention is now being focused on
rare variants that individually have small effects. In response,
numerous recent papers have proposed testing strategies to assess
association between a group of rare variants and a trait, with
competing claims about the performance of various tests. The power of a
given test in fact depends on the nature of any association and on the
rareness of the variants in question. We review such tests within a
general framework that covers a wide range of genetic models and types
of data. We study the performance of specific tests through exact or
asymptotic power formulas and through novel simulation studies of over
10,000 different models. The tests considered are also applied to real
sequence data from the 1000 Genomes project and provided by the GAW17.
We recommend a testing strategy, but our results show that power to
detect association in plausible genetic scenarios is low for studies of
medium size unless a high proportion of the chosen variants are causal.
Consequently, considerable attention must be given to relevant
biological information that can guide the selection of variants for testing.
\end{abstract}

% KEYWORDS

\end{frontmatter}

%s1 #&#
\section{Introduction}
\label{sec1}

Genome-wide association studies (GWAS) have identified numerous genetic
variants (single
nucleotide polymorphisms, or SNPs) that are associated with complex
human traits [e.g., \citet{Manolioetal2008}, \citet
{Hindorffetal}]. However, because of their limited sample sizes, such
studies are effective only at identifying common variants, that is, for
which the minor allele frequency (MAF) is not too small (e.g., MAF
$\geq$5\% for sample size $\sim$2000). In addition, variants that
have been identified through GWAS explain only small fractions of the
estimated trait heritabilities.
There is now much interest in understanding the role of rare variants
(as represented by SNPs with small MAFs), but because they are rare it
is difficult to detect associations with specific traits [e.g., \citet
{Bansaletal}; \citet{AsimitandZeggini}]. Next generation sequencing
(NGS) can produce detailed information on rare variants but studies
involving large numbers of individuals are not yet practical due to
cost, heterogeneity and other concerns. Attention has consequently
focused on methods that combine information across multiple rare SNPs
in a genomic region (see Section~\ref{sec6} for discussion on the practical
choice of a genomic region and SNPs within the region for analysis and
its impact on the statistical inference). This area is the focus of our
article. Our purpose is to review methods of testing for association
between rare variants and a trait, unify the different methods, and
give some new results.

To motivate our discussion, we refer to data from the Genetic Analysis
Workshop 17 (GAW 17) [\citet{Almasyetal}, 1000 Genomes Project Consortium
(\citeyear{1000GenomesProjectConsortium})]. These data include real sequence data
(SNP genotypes) obtained from the 1000 Genomes Project, and simulated
phenotype data (trait values) simulated by the GAW 17 committee. We
focus here on a single quantitative trait, Q2. The values of Q2 and
other traits were simulated for each person using normal linear
regression models that included the SNP effects and, in some cases,
additional covariates. Details concerning the simulation of trait
values are given by \citet{Almasyetal}. For Q2 the regression model
involved effects for 72 SNPs within 13 genes, with MAFs ranging from
0.07\% to 17.07\%. Our objective is to look for evidence of
associations between rare variants and Q2.

Papers that propose pooled association testing strategies for rare
variants include \citet{MorgenthalerandThilly}, \citet{LiandLeal},
\citet{MadsenandBrowning}, \citet{Bansaletal}, \citet{HanandPan}, \citet
{Hoffmannetal}, \citet{MorrisandZeggini}, \citet{Priceetal}, \citet
{YiandZhi}, \citet{Nealeetal}, \citet{Wuetal}, \citet{Su}  and \citet{Leeetal}. This
previous work has provided many tests but insight into settings when a
method will perform well, indifferently or poorly is still limited.
Recently, \citet{BasuandPan} and \citet{Ladouceur} conducted
extensive\vadjust{\goodbreak}
empirical evaluation (simulation) studies and reached a similar
conclusion that \textit{``the power of recently proposed statistical
methods depend strongly on the underlying hypotheses concerning the
relationship of phenotypes with each of these three factors''}:
proportions of causal variants, directions of the associations
(deleterious, protective or both), and the relationship between variant
frequencies and genetic effects [\citet{Ladouceur}]. However, the joint
effects of these factors have not been quantified analytically.
Moreover, the test procedures assume that SNPs have been placed in
groups, with pooling and testing carried out for SNPs within a given
group. There are various ways SNPs might be grouped and this will
affect the three factors mentioned. Ways of grouping SNPs are currently
being studied in connection with the recent Genetic Analysis Workshop
18 (GAW 18) and elsewhere.

In this paper we consider tests for genotype--phenotype association
within a unified framework. Most existing test statistics are either
linear statistics that are powerful against specific association
alternatives [e.g., \citet{MorgenthalerandThilly}, \citet{LiandLeal},
\citet{MorrisandZeggini}, \citet{MadsenandBrowning} and \citet
{Priceetal}] or quadratic statistics that have reasonable power across
a wide range of alternatives [e.g., \citet{Nealeetal}, \citet{Wuetal},
\citet{Leeetal}].
We study both classes of statistics theoretically and empirically and
provide several new insights. In particular, we examine the (asymptotic
or exact) powers of various tests as a function of the three factors
above. We deal with both categorical and quantitative traits, and allow
trait-dependent selection of individuals in a study as well as
nonindependent SNPs. We conduct novel simulation studies that
complement other recent empirical investigations and shed new light on
methods' comparison. We also discuss so-called optimality of tests and
indicate what this means in practical settings.

A feature of many of the linear statistics and of the quadratic
statistics of \citet{Wuetal} and \citet{Leeetal} is the use of weights
associated with individual SNPs, because of the suggestion that rarer
variants tend to have larger genetic effects. We demonstrate that even
if this assumption is true, using weights inversely proportional to
MAFs can in some cases have an adverse effect. We also show that for
linear statistics, methods of weight selection based on estimated
effects [e.g., \citet{HanandPan}, \citet{YiandZhi}, \citet
{Hoffmannetal}, \citet{LinandTang}] are similar to using quadratic statistics.

A referee has stressed the importance of several caveats concerning
the type of data considered in the paper, and hence the ``success'' of
testing procedures such as discussed here. First, errors in sequencing
data commonly occur. Methods for addressing this have not yet been well
studied in the present context, and we assume that genotypes are as
given. Methods used in other contexts [\citet{Dayeetal.}, \citet
{Skotteetal.}] are typically based on estimated sequencing error
probabilities, but we note that their accuracy is not well established
in specific settings. A second caveat is that the identification of
rare variants is difficult because of their low frequency, and because
sequencing errors can substantially affect the estimation of small
MAFs. They can also lead to a SNP that is actually monomorphic being
identified as a rare polymorphic SNP in some instances. Finally, the
nature and level of heritability explained by rare variants is at this
point speculative and it is unclear whether major successes will occur
from the approaches considered here. We take pains in the paper to
consider a broad range of genetic models but we cannot of course answer
questions about the scientific fundamentals.

%{\bf[To be revised]}
The remainder of the paper is organized as follows. Section~\ref{sec2}
introduces the framework for testing the association between a group of
rare variants and a general trait, reviews tests that have been
proposed along with analytical results relating the power of linear and
quadratic statistics to the various factors, and considers adjustment
for covariates. Section~\ref{sec3} presents theoretical power calculations for
normally distributed traits that
clarify when various methods will do well and the effects of using
weights. Section~\ref{sec4} gives numerical results based on large-scale
simulation studies of over 10,000 different models for both
quantitative and binary traits. Section~\ref{sec5} examines the GAW17
quantitative trait Q2 and sequence data from the 1000 Genomes Project.
Section~\ref{sec6} concludes with some recommendations for pooled testing.
Online supplementary materials [\citet{Supp}] include details specific
about test statistics and additional tables and figures for the power
comparison studies.

%s2 #&#
\section{Score Tests for Association}\label{sec2}

%s2.1 #&#
\subsection{No Covariate Adjustment}\label{sec2.1}

We assume that a group of $J$ SNPs and a trait $Y$ are under
consideration. The objective is to test whether there is association
between $Y$ and one or more of the SNPs.
For a set of $n$ unrelated individuals, let $Y_{i}$ be the measured
trait value for individual $i$ and
$\mathbf Y =(Y_1, \ldots, Y_n)'$. Let $X_{i j}$ denote the SNP genotype for
individual $i$, $i = 1, \ldots, n$ and $j = 1, \ldots, J$; for
simplicity we assume that $X_{i j}$ denotes whether the rare allele is
present $(X_{i j } = 1)$ or absent $(X_{i j} = 0)$ and let $\mathbf
X_i=(X_{i1},\ldots,X_{iJ})'$. It is straightforward to consider the case
where $X_{i j}$ is the number of copies (0, 1 or 2) of the rare allele
for SNP ${j}$, but there will be no or very few individuals with two
rare alleles in a study of current typical size. We assume for now that
there is no adjustment for covariates, since many papers address only
this case. However, covariate adjustment is often important and we
consider it in Section~\ref{sec2.4}.

Our interoest is in testing the null hypothesis
%
%e2.1 #&#
\begin{equation}
H_{0}\dvtx \mathbf Y \mbox{ and }\mathbf X \mbox{ are independent}.
\label{eq:H0}
\end{equation}
Most proposed methods for testing $H_0$ are based on statistics that
are (weighted) linear or quadratic combinations of statistics $S_j$
which measure association between $Y$ and SNP $j$, $j=1,\ldots, J$.
Without loss of generality, we assume that $S_{j}$ is such that under
the null $E[S_{j}] = 0$ and $\operatorname{Var}(S_{j}) = \sigma^{2}_{0 j}$, and under
alternatives $E[S_j]=\mu_j$ and $\operatorname{Var}(S_j)=\sigma^2_j$.
To facilitate further discussion, we assume that $Y$ is defined so that
a SNP with $\mu_{j} > 0$ is termed deleterious, with $\mu_{j} < 0$ is
protective, and with $\mu_{j} = 0$ is neutral; both deleterious and
protective SNPs are causal variants. Let $\mathbf S = (S_1,\ldots, S_J)'$
and $E[\mathbf S]=\bolds\mu= (\mu_1,\ldots, \mu_J)'$, and assume for
simplicity that the hypothesis of no association (\ref{eq:H0}) is
equivalent to the null hypothesis
%
%e2.2 #&#
\begin{equation}
H_{0}\dvtx \bolds\mu= \mathbf0. \label{eq:H0.mu}
\end{equation}

There are various options for $S_{j}$, but the approaches referred to
in Section~\ref{sec1} can almost all be expressed in terms of statistics of the form
%
%e2.3 #&#
\begin{equation}
S_{j} = \sum^{n}_{i = 1}
(Y_i - \overline{Y})X_{i j}, \quad j = 1, \ldots, J,
\label{eq:Sj}
\end{equation}
where $\overline{Y} = \sum^{n}_{i = 1} Y_i/n$ [e.g., see \citet
{LinandTang}; \citet{BasuandPan}]. The $S_j$ arise as score statistics
in regression models for the two important cases where $Y_i$ is
normally distributed and binary, respectively. They also arise from
Poisson models for counts and for other models in the linear
exponential family [e.g., \citet{Leeetal}]. For completeness, we outline
this for the binary case in the supplementary materials [\citet{Supp}].
Other statistics, for example, Wald or likelihood ratio statistics,
could be used (see Section~\ref{sec2.4}), but score statistics are almost
universally used in this area, and we focus on them. We note that the
score statistics have the advantage of requiring only estimates
obtained under the null hypothesis. In some contexts it is also useful
to replace $Y_i - \overline{Y}$ in (\ref{eq:Sj}) with some other
function $\alpha_i$ of either $Y_{i}$ or its rank, with $\sum^{n}_{i =
1} \alpha_{i} = 0$. It should be noted that genotypes $X_{ij}$,
$j=1,\ldots,J$, are not assumed to be mutually independent in the
subsequent development.

%t1 #&#
\begin{table*}
\caption{Summary of different association tests for
analyzing rare variants.
This is not an exhaustive list of all existing tests (see Sections \protect\ref{sec2}
and \protect\ref{sec6} for additional examples). Tests derived from random effect models
and adaptive linear models are operationally similar to quadratic tests
(see Section \protect\ref{sec2.3} for discussion). Details of the notation: see Section
\protect\ref{sec2.1}. Briefly, $\mathbf S=(S_1,\ldots,S_J)'$ is a vector of test statistics for
a group of $J$ rare variants, $\mathbf w=(w_1,\ldots,w_J)'$ is a vector of
weights, $A$ is a positive definite (or semi-definite) symmetric
matrix, $\Sigma_0$ is a known or estimated covariance matrix for $\mathbf
S$, $p_j$ is the minor allele frequency (MAF) of SNP $j$, $f(p_j)
=1/\sqrt{p_j(1-p_j)}$ in Weighted-sum of Madsen and Browning
(\citeyear{MadsenandBrowning}),
$f(p_j)$ depends on the MAF via a $\mathit{Beta}$ distribution in SKAT of
Wu et~al. (\citeyear{Wuetal}), and $p_L$ and $p_Q$ are the
$p$-values from chosen Linear and
Quadratic tests}\label{tab:new}
\begin{tabular*}{\textwidth}{@{\extracolsep{\fill}}p{5cm}p{5cm}p{6cm}@{}}
\hline
\multicolumn{3}{c}{\textbf{Class of tests}} \\
\hline
\multicolumn{1}{c}{\textbf{Linear}} &\multicolumn{1}{c}{\textbf{Quadratic}} &
\multicolumn{1}{c@{}}{\textbf{Combined/Hybrid}} \\
\hline
$W_L = \mathbf w' \mathbf S$ & $W_Q=\mathbf S' A \mathbf S$ & $H(W_L, W_Q)$\\
\hline
\multicolumn{3}{c}{\textbf{Example of specific tests}} \\ \hline
$\mathbf w=\mathbf1$ (CAST, $W_{L1}$) & $A=I$ (SSU and C-alpha, $W_C$)&
$\operatorname{max}_{w} \{W_L\}$ (EREC) \\
\citet{MorgenthalerandThilly} & \citet{Pan}, \citet{Nealeetal} & \citet
{LinandTang} \\[3pt]
$w_j=f(p_j)$ (Weighted-sum, $W_{Lp}$) & $A=\operatorname{diag}\{a_j\}$, $a_j=f(p_j)$
(SKAT) & $\operatorname{max}_{\rho\in[0,1]} (\rho W_L + (1-\rho) W_Q)$ (SKAT-O) \\
\citet{MadsenandBrowning} & \citet{Wuetal} & \citet{Leeetal} \\ [3pt]
$w_j=0$ if $p_j>$ threshold (Threshold) &$A=\Sigma_0^{-1}$ (Hotelling,
$W_H$) &$-2\log(p_L)-2\log(p_Q)$ (Fisher's method), $\min(p_L, p_Q)$
(minimum-$p$) \\
\citet{Priceetal} & \citet{BasuandPan} & \citet{derkach} \\
\hline
\end{tabular*}
\end{table*}

Many authors have considered linear test statistics for $H_0$ (\ref
{eq:H0.mu}) of the form
%
%e2.4 #&#
\begin{equation}
W_{L} = \sum^{J}_{j = 1}
w_{j} S_{j}={\mathbf w'\mathbf S}, \label{eq:WL}
\end{equation}
where the weights $w_{j}$s are specified nonnegative values and $\mathbf
w=(w_1,\ldots,w_J)'$. \citet{BasuandPan} provided a review, and we note
two important cases: \citet{MorgenthalerandThilly} considered the
``cohort allelic sums test'' (CAST) where each $w_{j}=1$, and \citet
{MadsenandBrowning} based $w_{j}$ on the (estimated) MAF, with larger
weights for SNPs with smaller MAF. The rationale for the latter weights
is that causative SNPs would be subject to ``purifying selection'' and
so be rarer in the population than neutral SNPs, but evidence for this
so far seems slight. We also note that because the MAFs have to be
estimated, sequencing errors as discussed in Section~\ref{sec1} can have an
effect; we assume (idealistically) that such errors have not occurred.
\citet{Priceetal} also considered ``threshold'' versions in which $w_{j}
> 0$ only if the estimated MAF is below a specified threshold (e.g., 1\% or 5\%). Such linear composite statistics can have good power
against association alternatives where $\mu_j \geq0$, with $\mu_j > 0$
for some subset of $\{j = 1, \ldots, J \}$. However, their power may be
poor for alternatives where both positive and negative values of $\mu
_j$ are possible, and when only a small proportion of the $J$ SNPs are
causal and have $\mu_j>0$ [\citet{Nealeetal}, \citet{BasuandPan}]. The
effects of association direction on different statistics are studied in
Sections~\ref{sec3} and \ref{sec4}.

Many authors have also considered quadratic statistics,
%
%e2.5 #&#
\begin{equation}
W_{Q} = \mathbf{S}^{\prime} A \mathbf{S}, \label{eq:WQ}
\end{equation}
where $A$ is a positive definite (or semi-definite) symmetric matrix.
One common choice is $A=\Sigma^{-1}_0$, where $\Sigma_0$ is a known or
estimated covariance matrix for $\mathbf S$ under $H_0$; this gives a
Hotelling statistic,
%
%e2.6 #&#
\begin{equation}
W_{H} = \mathbf{S}^{\prime} \Sigma^{-1}_0
\mathbf{S}. \label{eq:WH}\vadjust{\goodbreak}
\end{equation}
Other quadratic statistics include the ``SSU'' statistic of \citet{Pan}
and the ``C-alpha'' statistic of \citet{Nealeetal} which are based on
$A=I$, the $J \times J$ identity matrix; the ``SKAT'' statistic of \citet
{Wuetal} uses $A = \operatorname{diag}\{a_1,\ldots,a_J\}$, where the $a_j$s are
weights that depend on the MAFs via a $\mathit{Beta}$ function. The linear
statistic $W_L$ in (\ref{eq:WL}) can also be expressed in quadratic
form, since $W^2_L$ is equivalent to (\ref{eq:WQ}) with $A=\mathbf w\mathbf
w'$. However, note that $A$ is no longer positive definite in this
case. Quadratic statistics arise naturally from regression models
relating $Y$ and $X_j$ as we discus below. Finally, we remark that
recent work has considered combining evidence from linear and quadratic
statistics [e.g., \citet{Leeetal} and \citet{derkach}]. We discuss this
in Section~\ref{sec6}, but focus on individual linear and quadratic statistics
here ({Table~\ref{tab:new}}).

%s2.2 #&#
\subsection{Distributions of Linear and Quadratic Statistics Under
Normality}\label{sec:2.2}

It is instructive to consider the case where $\mathbf S$ is normally
distributed. For both binary and quantitative traits, the vectors $\mathbf
S$ are all at least asymptotically normal, and analytical derivations
of power and discussions of optimality rely on this assumption [e.g., \citet{LinandTang}; \citet{Leeetal}]. The case where $\mathbf S$ is normal
in finite samples also is well known in connection with tests for a
multivariate normal mean~$\bolds\mu$; see, for example, \citet{Mardia},
Chapter~5.
%For tests of (\ref{eq:H0.mu}), $H_0$: $\bolds\mu= \bm0$, the power of
%statistics such as as (\ref{eq:WL})-(\ref{eq:WH}) against an
%alternative hypothesis $H_1$ for which $\bolds\mu\neq\bm0$ depends on
%$\bolds\mu$ and on the distribution of $\bm S$ under $H_1$.

Suppose that under $H_1$ for which $\bolds\mu\neq\mathbf0$ the
distribution of $\mathbf S$ is (exactly or asymptotically) multivariate
normal with mean $\bolds\mu$ and covariance matrix $\Sigma$, $\mathbf S \sim
N(\bolds\mu, \Sigma)$. For simplicity we assume that $\Sigma$ is known;
this is allowable for asymptotic results which we focus on here. In
finite samples where $Y$ given $X$ is normal, the effect of estimating
$\Sigma$ is to replace normal and chi-square distributions below with
$t$ and $F$ distributions, respectively. With $J$ fixed and $n$ going to
infinity, these converge to the normal and chi-square distributions we consider.

Let $\lambda_1,\ldots, \lambda_J$ be the eigenvalues of $\Sigma
^{1/2}A\Sigma^{1/2}$ and $P$ be the $J \times J$ orthogonal matrix
whose columns are the corresponding eigenvectors. Then the following
distributional results hold [e.g., \citet{Rao}, Section 3b.4]:
\begin{longlist}[(iii)]
\item[(i)] $W_Q$ is distributed as a linear combination of independent
noncentral $\chi^2_1$ random variables,
%
%e2.7 #&#
\begin{equation}
W_{Q} \sim\sum_{j=1}^{J}
\lambda_j \chi^2_{1, \mathrm{nc}_j} \label{eq: 2.7},
\end{equation}
where $\chi^2_{k, r}$ denotes a noncentral $\chi^2$ random variable
with $k$ degrees of freedom and noncentrality parameter $r$, and $\mathrm{nc}_j
= (\{P'\Sigma^{-1/2}\bolds{\mu}\}_j)^2$.
\item[(ii)] If $A=\Sigma^{-1}$, then $W_Q \sim\chi^2_{J,\mathrm{nc}}$ with $\mathrm{nc}
= \bolds\mu' \Sigma^{-1} \bolds\mu$. If $\Sigma= \Sigma_0$, then $W_Q$ is
the Hotelling statistic~(\ref{eq:WH}).
\item[(iii)] $Z^2_L=W^2_L/(\mathbf w' \Sigma\mathbf w)=(\mathbf w' \mathbf S)^2/(\mathbf
w' \Sigma\mathbf w) \sim\chi^2_{1,\mathrm{nc}}$ with $\mathrm{nc} = (\mathbf w' \bolds\mu)^2 /
(\mathbf w' \Sigma\mathbf w)$ when $\Sigma = \Sigma_0$. When this is not
true, then the distribution of $Z^2$ is a multiple of the noncentral
$\chi^2_1$ random variable.
\item[(iv)] Under the null hypothesis $H_0$: $\bolds\mu=\mathbf0$,
$W^2_L/\break  (\mathbf w' \Sigma_0 \mathbf w)$ is a $\chi^2_1$ random variable;
$W_Q$ is a linear combination of independent $\chi^2_1$ random
variables with each $\mathrm{nc}_j=0$ in (\ref{eq: 2.7}).
\end{longlist}

It should be noted that no adjustment is needed to reflect the fact
that $\mathbf w$ may involve estimated MAFs. This is because the
distributional results are based on the sampling distribution of $Y$
given $X_{ij}$, where estimates of MAFs are functions of $X$ alone and
so are treated as fixed in this section. We return to this point in
Section~\ref{sec4.1}, and we also note in Section~\ref{sec4.2} that complications\vadjust{\goodbreak} arise
when retrospective (case--control) studies are used with binary responses.
%{\bf To be revised: what is the effect? $\beta$ or $EV$? need to be
%clear}
These results allow the power against a simple alternative hypothesis
$H_1$ with a specified $\bolds\mu\neq\mathbf0$ to be calculated for any
linear test statistic (\ref{eq:WL}) or quadratic test statistic (\ref
{eq:WQ}). Critical values for a test of $H_0$: $\bolds\mu=\mathbf0$ are
obtained according to (iv). Software exists for the computation of
probabilities associated with linear combinations of central or
noncentral $\chi^2_1$ random variables, for example, the \mbox{\textit
{CompQuadForm}} package in R [\citet{DuchesneanddeMicheaux}]. In
particular, we note that:

\begin{longlist}[(a)]
\item[(a)] For a size $\alpha$ test using the linear statistic $W_L$ in
(\ref{eq:WL}) or, equivalently, $Z_L^2$ in (iii) above, the $\alpha$
critical value is $\chi^2_{1}(1-\alpha)$, the $1-\alpha$ quantile for
the $\chi^2_1$ distribution. (The test is two-sided to allow for either
positive or negative $W_L$ under $H_1$.) The power against $H_1$ when
$\Sigma= \Sigma_0$ is
%
%e2.8 #&#
\begin{eqnarray}\label{eq:powerL}
P\bigl(\chi^2_{1,\mathrm{nc}_L}>\chi^2_1(1-
\alpha)\bigr)
\nonumber
\\[-8pt]
\\[-8pt]
\eqntext{\mbox{where } \mathrm{nc}_L=\bigl(\mathbf w' \bm \mu
\bigr)^2/\bigl(\mathbf w' \Sigma\mathbf w\bigr). }
\end{eqnarray}
\item[(b)] For a size $\alpha$ test using the Hotelling statistic $W_H$
in (\ref{eq:WH}), the $\alpha$ critical value is $\chi^2_{J}(1-\alpha
)$. The power against $H_1$ in the case where $\Sigma=\Sigma_0$ is
%
%e2.9 #&#
\begin{eqnarray}\label{eq:powerQ}
P\bigl(\chi^2_{J,\mathrm{nc}_H}>\chi^2_J(1-
\alpha)\bigr)
\nonumber
\\[-8pt]
\\[-8pt]
\eqntext{\mbox{where } \mathrm{nc}_H = \bolds\mu '
\Sigma^{-1}\bolds\mu. }
\end{eqnarray}
\end{longlist}
The specific power of both statistics depends on $\bolds\mu$ and on the
distribution of $\mathbf S$ under $H_1$, however, some general features can
be seen. For simplicity, suppose $\Sigma= \Sigma_0$ and that $\Sigma$
is diagonal (SNPs are independent).
The quadratic statistic $W_H$ (\ref{eq:WH}) is a reasonable choice when
both deleterious $(\mu_j>0)$ and protective $(\mu_j<0)$ SNPs are
plausible, because $\mathrm{nc}_H$ is a function of the $\mu_j^2$. The statistic
$W_H$ can be decomposed as
$W_H = Z^2_L + R$,
where $Z_L$ and $R$ are independent under $H_1$, and $R\sim\chi
^2_{J-1,\mathrm{nc}_R}$ with
$\mathrm{nc}_R = \mathrm{nc}_H - \mathrm{nc}_L = \bolds\mu' \Sigma^{-1} \bolds\mu- (\mathbf w' \bolds\mu
)^2/(\mathbf w' \Sigma\mathbf w)$.
The linear statistic $W_L$ is optimal when $\mathrm{nc}_R = 0$, but
the advantage of $W_L$ over the quadratic statistic $W_H$ disappears
as $\mathrm{nc}_R$ increases. We will discuss this in Sections~\ref{sec3} and \ref{sec4}.

%s2.3 #&#
\subsection{Additional Considerations: Optimality, Random Effect
Models, Adaptive Linear Models, $p$-Values and Permutation Distribution}\label{sec2.3}

A number of authors [e.g., \citet{Leeetal}, \citet{Nealeetal}, \citet
{LinandTang}] have claimed to obtain ``optimal'' tests. This is
theoretically possible if we specify a suitable family of test
statistics, but for this to be of practical use we must have strong
prior knowledge about the alternative hypothesis. For example, among
the class of linear statistics (\ref{eq:WL}),\vadjust{\goodbreak} maximal power is obtained
when $\mathbf w = \Sigma^{-1} \bolds\mu$. When the $S_j$s are independent so
that $\Sigma= \operatorname{diag}\{\sigma^2_1,\ldots,\sigma^2_J\}$, this gives $w_j=\mu
_j/\sigma^2_j$. This linear statistic is (asymptotically) optimal among
all tests of fixed size based on $\mathbf S$, assuming $\bolds\mu$ is known.
Quadratic statistics (\ref{eq:WQ}) for which $A$ has rank 2 or more can
never be optimal against a specific alternative $(\bolds\mu, \Sigma)$.
However, quadratic tests can maintain reasonable power over wide ranges
of alternatives, whereas a linear statistic's power can be poor except
near a specific alternative. \citet{Goemanetal} and other authors have
discussed optimality of score statistics coming from random effects
models, but these results are also based on averaging over a family of
alternatives, which may or may not be plausible in a given setting. For
example, quadratic statistics (\ref{eq:WQ}) can be obtained from random
effect regression models in which $Y$ is related to $\mathbf X$ through a
linear function $\bolds\beta' \mathbf X$ and the $J \times1$ regression
coefficient $\bolds\beta$ is a random vector with mean $\mathbf0$ and
covariance matrix $\tau A$. The hypothesis $\tau= 0$ then corresponds
to $H_0$ in (\ref{eq:H0}) and a score statistic for testing it is
[\citet{Goemanetal}, \citet{BasuandPan}]
%
%e2.10 #&#
\begin{equation}
W'_Q = \tfrac{1}{2}\mathbf S' A \mathbf S -
\tfrac{1}{2}\operatorname{trace}(A\Sigma_0). \label{eq:WQ'}
\end{equation}
Using $W'_Q$ is equivalent to using $W_Q$ in (\ref{eq:WQ}) when $\Sigma
_0$ is known. The first term in (2.10) also arises from other score
tests in generalized linear models [\citet{Leeetal}]. In general,
$\Sigma_0$ (and $A$) involve estimates and asymptotic distributions for
$W_Q$ are used to get $p$-values. The asymptotic distributions are
typically of the form (\ref{eq: 2.7}), but with the $\lambda_j$
involving estimates. We comment further on the calculation of $p$-values
at the end of this section.

Some authors [e.g.,  \citet{HanandPan}, \citet{Hoffmannetal}, \citet
{LinandTang}] have proposed two-stage or other adaptive approaches in
which the weighting vector ${\mathbf w}$ for $W_L$ in (\ref{eq:WL}) is
chosen after preliminary examination of the direction of $S_{j}$ or an
estimate of its effect based on the observed data, in a hope of
choosing an ``optimal'' weight. However, such an approach cannot on its
own (i.e., without the use of additional information from other
sources) improve globally the linear statistics. In fact, if we choose
the $\mathbf w$ that maximizes the standardized linear test statistic~(\ref
{eq:WL}), then we end up with the quadratic statistic~(\ref{eq:WH}). In
particular [e.g., \citet{Mardia}, page 127, or \citet{LiandLagakos},
Section~3],
\[
\mathop{\operatorname{sup}}_{\mathbf w} \biggl\{ \frac{W_L^2}{\operatorname{Var}(W_L)} \biggr\} =
\mathop{\operatorname{sup}}_{\mathbf w} \biggl\{ \frac{ ( {\mathbf w}^{\prime} {\mathbf S}
)^{2}}{{\mathbf w}^{\prime} \Sigma {\mathbf w}} \biggr\} = {\mathbf
S}^{\prime} \Sigma^{- 1} {\mathbf S}=W_H,
\]
where the maximizing vector is ${\mathbf w} = \Sigma^{- 1} {\mathbf S}$.
This helps explain why \citet{BasuandPan} found that adaptive
procedures did not perform as well as one might have hoped.

\citet{LinandTang} have proposed a test statistic $T_{{\mathrm{max}}}$ based on
the maximum of a specified set of $K$ linear statistics, each with
different weights, $T^2_k =(\mathbf w'_k \mathbf S)^2/(\mathbf w'_k\Sigma\mathbf
w_k)$. We do not consider such statistics here, but it is clear that
their performance will depend on the choice of ``appropriate'' weighting
vectors $\mathbf w_k$. When there is little prior information and the $\mathbf
w_k$s are selected to cover a wide range of alternatives, it seems
likely that $\operatorname{max}(T^2_k)$ would be similar to $W_H$. A similar
suggestion involving quadratic statistics is made by \citet{Leeetal}.
In practice, there is often very limited prior information about the
nature of $\bolds\mu$, especially concerning which SNPs might be causal,
so one cannot be confident that a linear test statistic will be
effective, nor which quadratic statistics might be the best. Sections~\ref{sec3}
and \ref{sec4} investigate situations in which specific statistics will be more powerful.

To achieve reasonable power, sample sizes have to be rather large, as
we discuss in Section~\ref{sec4}. The calculation of $p$-values, critical values
or power is often based on large sample approximations given by normal
and chi-square distributions in Section~\ref{sec:2.2}. In general, this requires
estimation of matrices $\Sigma_0$ and $A$ (as do test statistics
themselves) but with consistent estimators the limiting distributions
provide adequate approximation for sufficiently large samples. In
general, a consistent estimator of $\Sigma_0$ for $\mathbf S$ given by (\ref
{eq:Sj}) is
%
%e2.11 #&#
\begin{equation}
\hat{\Sigma}_0 =\frac{\sum_{i=1}^{n}(Y_i-\overline{Y})^2}{n-1}X'_c
X_c, \label{eq:sigmap}
\end{equation}
where $ X'_c$ has $(i,j)$ entry $X_{ij} - \overline{X}_j$ (where
$\overline{X}_j = \sum_{i=1}^{n}X_{ij}/n$). However, because events
with $X_{ij}=1$ are rare, the distribution of $\mathbf S$ can be quite
nonnormal even in rather large samples, and more accurate ways to
calculate $p$-values and critical values are needed, especially for
quadratic statistics. Some authors [e.g., \citet{Leeetal}] have given
skewness or kurtosis adjustments that seem to improve accuracy in
certain settings. More generally, however, we can obtain $p$-values (and
study power) by simulation. When there is no adjustment for covariates,
the permutation distribution of $\mathbf S=(S_1,\ldots,S_J)'$ is typically
used [e.g., \citet{BasuandPan}]; this is the distribution that arises
from randomly permuting the $Y_i$s and assigning them to the $\mathbf
X_i$s. This also applies when $Y$ is a discrete variable, when $X_{i
j}$s are correlated within individuals (e.g., due to linkage
disequilibrium, LD) and when sampling of the individuals is
$Y$-dependent. More generally, when there are covariates present, we
may need to rely on bootstrap simulations. We comment on this in the
following section.
%s2.4 #&#
\subsection{Adjustment for Covariates}\label{sec2.4}

\citet{LinandTang} and \citet{Wuetal} have stressed that adjustment for
covariates and population stratification will be important in many
contexts involving rare variants. In this case we use regression
models; for illustration, we consider the case of a binary trait.
Suppose that in addition to the genotype vector ${\mathbf X}_{i}$ there is
a vector ${\mathbf v}_{i}$ of covariates that may be related to a binary
trait $Y_{i}$. Then a logistic regression model
%
%e2.12 #&#
\begin{eqnarray}\label{eq: 7.1}
&&\Pr (Y_{i} = 1 | {\mathbf X}_{i}, {\mathbf v}_{i} )
\nonumber
\\[-8pt]
\\[-8pt]
\nonumber
&&\quad=
\frac{\exp(
\beta_0 + {\bolds\beta}^{\prime} {\mathbf X}_{i} + {\bolds\gamma}^{\prime} {\mathbf
v}_{i})}{1 + \exp( {\beta_0} + {\bolds\beta}^{\prime} {\mathbf X}_{i} + {\bm
\gamma}^{\prime} {\mathbf v}_{i})}= {\mu_i}
\end{eqnarray}
might be considered, and a test of $H_{0} \dvtx {\bolds\beta} = {\mathbf0}$ can
be carried out. For testing rare variants some authors have replaced
the term ${\bolds\beta}^{\prime} {\mathbf X}_{i}$ in (\ref{eq: 7.1}) with
$\beta r_{i}$, where $r_{i} = \sum^{J}_{j = 1} X_{i j}$ is the total
number of rare variants per individual [e.g., \citet
{MorrisandZeggini}; \citet{YilmazandBull}], but this corresponds to
using a linear statistic in previous sections and can be ineffective.
We consider the case where ${\bolds\beta} = (\beta_{1}, \ldots, \beta
_{J})^{\prime}$ in order to examine settings for which causal SNPs may
be either deleterious or beneficial. Consideration of the power of
alternative tests in large samples parallels the discussion in Section~\ref{sec:2.2}, as follows.

Let $\hat{\bolds\beta}$ be the estimator of ${\bolds\beta}$ based on the
model in question and assume that under $H_{0}\dvtx {\bolds\beta} = {\mathbf0}$,
the asymptotic distribution of $\sqrt{n} {\hat{\bolds\beta}}$ is
multivariate normal with mean ${\mathbf0}$ and covariance matrix $\Sigma$.
Following \citet{LiandLagakos}, we consider a sequence of contiguous
alternatives
%
%e2.13 #&#
\begin{equation}
H^{(n)}_{1} \dvtx {\bolds{\beta}} = {\mathbf b} / \sqrt{n},
\label{eq: 7.2}
\end{equation}
where ${\mathbf b} = (b_{1}, \ldots, b_{J})^{\prime}$ is a specified
vector. Under this sequence as $n \rightarrow\infty$ the distribution
of $\sqrt{n} {\hat{\bolds\beta}}$ approaches a multivariate normal
distribution with mean ${\mathbf b}$ and covariance matrix $\Sigma$. Thus,
asymptotic power for a test statistic can be computed in the same way
as in Section~\ref{sec2.3}. \citet{ LiandLagakos} compare the quadratic Wald
test statistic $W = \hat{\bolds\beta}{}^{\prime} \hat{\Sigma}^{- 1} \hat
{\bolds\beta} $, where $\hat{\Sigma}$ is a consistent estimate of $\Sigma $ under $H_{0}$, with linear statistics $Z = {\mathbf a}^{\prime}
\hat{\mathbf
\beta}$. These are analogous to (\ref{eq:WH}) and (\ref{eq:WL}),
respectively. The likelihood score statistic for testing $\bolds\beta=\mathbf
0$ is an alternative to the Wald statistic; it is easily found as
[e.g., \citet{LinandTang}]
%
%e2.14 #&#
\begin{equation}
\mathbf U = \sum_{i=1}^{n}(Y_i-
\hat{\mu}_i)\mathbf X_i, \label{eq: 7.3}
\end{equation}
where $\hat{\mu}_i=e^{\hat{\beta}_0+\hat{\bolds\gamma}'\mathbf
v_i}/(1+e^{{\hat{\beta}_0}+\hat{\bolds\gamma}'\mathbf v_i})$ and ${\hat{\beta
}_0}$, $\hat{\bolds\gamma}$ are estimated from (\ref{eq: 7.1}) when $\mathbf
\beta=\mathbf0$. It also follows from standard maximum likelihood large
sample theory that the covariance matrix of $\mathbf U$ under $H_0$ is
estimated consistently by
%
%e2.15 #&#
\begin{eqnarray}\label{eq: 7.4}
\hat{\Sigma}_{U}= \widehat{\operatorname{Var}}(\mathbf U)& =& \Biggl(\sum
_{i=1}^{n}\hat {\sigma}^2_i
\mathbf X_i\mathbf X'_i \Biggr) \nonumber\\
&&{}- \Biggl(\sum
_{i=1}^{n}\hat{\sigma }^2_i
\mathbf X_i\tilde{\mathbf v}'_i \Biggr) \Biggl(\sum
_{i=1}^{n}\hat{\sigma }^2_i
\tilde{\mathbf v}_i\tilde{\mathbf v}'_i
\Biggr)^{-1} \\
&&{}\cdot\Biggl(\sum_{i=1}^{n}
\hat {\sigma}^2_i\tilde{\mathbf v}_i\mathbf
X'_i \Biggr), \nonumber
\end{eqnarray}
where $\hat{\sigma}^2_i=\hat{\mu}_i(1-\hat{\mu}_i)$ and $\tilde{\mathbf
v}_i = (1,\mathbf{v}'_i)'$. These correspond to results given by \citet
{LinandTang}, who consider linear statistics based on linear
combinations of the elements $U_1,\ldots,U_J$ of $\mathbf U$. The statistic
(\ref{eq: 7.3}) and variance estimate (\ref{eq: 7.4}) are given here
for prospective sampling but can be shown to apply under case--control
sampling. As in Sections~\ref{sec2.1}--\ref{sec2.3}, test statistics such as
$W^{*}_{H}=\mathbf U'\hat{\Sigma}_{U}^{-1}\mathbf U$ and $W_L^{*} = (\mathbf w'
\mathbf U)/(\mathbf w'\hat{\Sigma}_{U}^{-1}\mathbf w)$, which correspond to
$W_H$ and $W_L$ in preceding sections, can be used. When there are no
covariates $\mathbf v_i$, it is readily seen that (\ref{eq: 7.3}) reduces
to (\ref{eq:Sj}) and that (\ref{eq: 7.4}) equals $(n-1)/n$ times (\ref
{eq:sigmap}). It should be noted that when covariates $\mathbf v_i$ are
present, the normal approximations considered earlier apply, but the
permutation distribution $p$-values do not unless the $\mathbf X_{i}$s are
independent of the $\mathbf v_i$. \citet{LinandTang} suggest a parametric
bootstrap as an alternative, based on randomly generating response
$Y_i$s from the fitted null model based on ${\hat{\beta}_0}$, $\hat{\bolds\gamma}$.

Normal linear regression models for quantitative variables $Y$ also
produce score statistics of the form (\ref{eq: 7.3}) with $\hat{\mu}_i
= \hat{\beta}_0+\hat{\gamma}'\mathbf v_i$, as do certain other generalized
linear models [\citet{Leeetal}]. It should be mentioned that in the
case of quantitative $Y$-dependent sampling and models with supplementary
covariates $\mathbf{v}_{i}$ as in (\ref{eq: 7.1}), adjustments to
estimating functions [e.g., \citet{Huangandlin}; \citet{YilmazandBull}]
are needed; this is beyond our present scope, but we note that
statistics like (\ref{eq: 7.3}) arise once again [\citet{Burnett}].

%s3 #&#
\section{Normally Distributed Traits}\label{sec3}

%s3.1 #&#
\subsection{Distributions of the Linear and Quadratic Statistics}\label
{sec:normalD}
To provide more insights on the effects of the choice of linear vs. quadratic statistics and the use of weights on power, it is helpful to
consider genetic scenarios described by a normal linear model,
%
%e3.1 #&#
\begin{eqnarray}\label{eq:model}
Y_i = \beta_0 + \beta_1X_{i1}+\cdots+
\beta_1X_{iJ} + e_i
\nonumber
\\[-8pt]
\\[-8pt]
\eqntext{\mbox{for } i=1, \ldots,
n,}
\end{eqnarray}
with $e_i \sim N(0,\sigma^2)$ and the $X_{ij}$s mutually independent
Bernoulli variables with $P(X_{ij}=1)=p_j$, approximately twice the MAF
of SNP $j$, $j=1,\ldots,J$.
The score statistic $\mathbf S=(S_1,\ldots,S_J)'$ with
%
%e3.2 #&#
\begin{equation}
S_j = \sum_{i=1}^{n}{(Y_i-
\overline{Y})X_{ij}} = \sum_{i=1}^{n}{(X_{ij}-
\overline{X}_j)Y_i} \label{eq:Sj2}
\end{equation}
arises from maximum likelihood theory for testing
$
H_0\dvtx \bolds\beta=(\beta_1,\ldots,\beta_J)'=\mathbf0
\label{eq:H0beta}
$, as noted in Section~\ref{sec2.4}. Normal models are widely used for
quantitative traits such as blood pressure or lipid levels. Due to the
normality of $Y$, the distribution of $S_j$ given the genotypes is
$S_j \sim N(m_j(1-m_j/n) \beta_j, m_j(1-m_j/n)\sigma^2)$, where $m_j =
\sum_{i=1}^n X_{ij}$. For any given sample the $m_j$ are treated as
fixed values, and for simplicity we consider the case where $m_j$ is
equal to its expected value $np_j$ so that
%
%e3.3 #&#
\begin{equation}
\mathbf S \sim N(\bolds\mu, \Sigma), \label{eq: 4.4}
\end{equation}
where
$\bolds\mu= (np_1(1-p_1)\beta_1, \ldots, np_J(1-p_J)\beta_J)' $
and
$ \Sigma=\operatorname{diag}\{np_1(1-p_1)\sigma^2, \ldots, np_J(1-p_J)\sigma^2\}$.
As earlier, we ignore the small effects due to the need to estimate
$\sigma^2$ in large samples.

Here and in simulations below, we consider settings according to the
variation of $Y$ explained by the set of SNPs. Under model (\ref{eq:model}),
the total phenotypic variation explained by the $J$ SNPs is
%
%e3.4 #&#
\begin{eqnarray}\label{eq:EV}
\mathit{EV}& = &\frac{\operatorname{Var}(E[Y|\mathbf X])}{\operatorname{Var}(Y)} = \frac{\sum_{j=1}^{J}p_j(1-p_j)\beta^2_j}{\sum_{j=1}^{J}p_j(1-p_j)\beta^2_j+\sigma
^2}\nonumber\\
&\approx&\sum
_{j=1}^{J} {p_j(1-p_j)
\beta_j^2}/{\sigma^2} \\
&= &\sum
_{j=1}^{J} \mathit{EV}_j, \nonumber
\end{eqnarray}
where $\mathit{EV}_j =p_j(1-p_j)\beta^2_j/\sigma^2$ is the ``Explained
Variation'' by SNP $j$. The approximation assumes that the phenotypic
variation explained by genetic factors is small, which is in agreement
with current data. The distribution of $W_L=\mathbf w' \mathbf S$ is
$N(n \sum_{j=1}^{J}{w_j p_j(1-p_j)\beta_j},  n \sum_{j=1}^{J}{w_j^2
p_j(1-p_j)\sigma^2})$,\break  and
%
%e3.5 #&#
\begin{equation}
W_L^2\Big/\Biggl(\sum_{j=1}^{J}{w_j^2
p_j(1-p_j)\sigma^2}\Biggr) \sim
\chi^2_{1, \mathrm{nc}_L}, \label{eq:WL2}
\end{equation}
where
%
%e3.6 #&#
\begin{eqnarray}\label{eq:ncL}\qquad
\mathrm{nc}_L &=& n \frac{(\sum_{j=1}^{J}w_jp_j(1-p_j)\beta_j/\sigma)^2}
{\sum_{j=1}^{J}w^2_jp_j(1-p_j)}
\nonumber
\\[-8pt]
\\[-8pt]
\nonumber
&=& n \frac{(\sum_{j=1}^{J} w_j \operatorname{sign}(\beta
_j)\sqrt{p_j(1-p_j)} \sqrt{\mathit{EV}_j})^2}{\sum_{j=1}^{J}w^2_jp_j(1-p_j)}.
\end{eqnarray}

Similarly, assuming $A = \operatorname{diag}\{a_1,\ldots, a_J\}$ where the $a_j$s can
also be interpreted as weights for quadratic statistics $W_Q=\mathbf S' A
\mathbf S$, we have
%
%e3.7 #&#
\begin{equation}
W_Q \sim \sum_{j=1}^{J}
\lambda_j\chi^2_{1, \mathrm{nc}_j}, \label{eq:WQ2}
\end{equation}
where
%
%e3.8 #&#
\begin{eqnarray}\label{eq:ncj}
\lambda_j& = & a_j np_j(1-p_j)
\sigma^2\quad\mbox{and }
\nonumber
\\[-8pt]
\\[-8pt]
\nonumber
\mathrm{nc}_{j}&=&np_j(1-p_j)
\beta^2_j/\sigma^2=n\mathit{EV}_j.
\end{eqnarray}

%s3.2 #&#
\subsection{Effects of Weights and Genetic Factors on Power}
\label{sec:normalsub}

We consider for discussion two linear statistics $W_L=\mathbf w' \mathbf S$: $W_{L1}$ with $w_j\equiv1$ [\citet{MorgenthalerandThilly}] and
$W_{Lp}$ with $w_j=1/\sqrt{p_j(1-p_j)}$ [\citet{MadsenandBrowning}]. We
also consider two quadratic statistics $W_Q=\mathbf S' A \mathbf S$: $W_C$ with
$A=I$ ($a_j\equiv1$) (C-alpha) and the Hotelling $W_{H}$ with $A=\Sigma
^{-1}$ ($a_j= 1/ (n p_j (1-p_j) \sigma^2)$). We note that the $p_j$ are
actually the values $\hat{p}_j=m_j/n$, but $\hat{p}_j=p_j$ here since
we are considering the situation where the values of $m_j$ are equal to
their expected values $np_j$. From (\ref{eq:WL2})--(\ref{eq:ncj}) we
then have
\[
W_{L1}^2\Big/\Biggl(\sum_{j=1}^{J}{p_j(1-p_j)
\sigma^2}\Biggr) \sim\chi^2_{1, \mathrm{nc}_{L1}},
\]
where
%
%e3.9 #&#
\begin{eqnarray}
\mathrm{nc}_{L1} &= & n \frac{(\sum_{j=1}^{J}p_j(1-p_j)\beta_j/\sigma)^2}
{\sum_{j=1}^{J}p_j(1-p_j)}
\nonumber
\\[-8pt]
\\[-8pt]
\nonumber
& =& n \frac{(\sum_{j=1}^{J} \operatorname{sign}(\beta_j)\sqrt {p_j(1-p_j)} \sqrt{\mathit{EV}_j})^2}{\sum_{j=1}^{J}p_j(1-p_j)}, \label{eq:ncL1}
\end{eqnarray}
\[
W_{Lp}^2/\bigl(J \sigma^2\bigr) \sim
\chi^2_{1, \mathrm{nc}_{Lp}},
\]
where
%
%e3.10 #&#
\begin{eqnarray} \label{eq:ncLp}
\mathrm{nc}_{Lp}& =& n \frac{(\sum_{j=1}^{J}\sqrt{p_j(1-p_j)}\beta_j/\sigma)^2}{J}
\nonumber
\\[-8pt]
\\[-8pt]
\nonumber
&=& n \frac{(\sum_{j=1}^{J} \operatorname{sign}(\beta_j) \sqrt{\mathit{EV}_j})^2}{J},
\\
W_{C} &\sim&\sum_{j=1}^{J}
\bigl(np_j(1-p_j)\sigma^2 \bigr)
\chi^2_{1, \mathrm{nc}_{j}},\nonumber
\end{eqnarray}
where $\mathrm{nc}_j =np_j(1-p_j)\beta^2_j/\sigma^2=n\mathit{EV}_j$ as in equation~(\ref
{eq:ncj}), and
\[
W_{H} \sim\chi^2_{J, \mathrm{nc}},
\]
where
$\mathrm{nc} = \sum_{j=1}^{J} \mathrm{nc}_j= n \sum^{J}_{j=1}\mathit{EV}_j \approx n
\mathit{EV}$.\vadjust{\goodbreak}

The above results show that the power of $W_H$ depends (approximately)
just on the total
explained variation $\mathit{EV}$ and sample size $n$, and it is not sensitive
to the direction of
the SNP effects [$\operatorname{sign}(\beta_j)$] nor the MAF $p_j$. Although the
C-alpha statistic $W_C$ uses ``equal'' weights for all SNPs, its power
depends not only on the $\mathit{EV}_j$s and $n$ but also on the $p_j$s, because
the corresponding coefficients for the linear combination of
independent $\chi_{1, \mathrm{nc}_j}^2$ are proportional to $p_j(1-p_j)$,
essentially giving
\textit{smaller weight to rarer variants}. The test statistic $W_C$ has
been found powerful in a wide range of settings for binary phenotypes
[e.g., \citet{Nealeetal}, \citet{BasuandPan}]. For the most part, the
settings investigated were ones where the regression coefficients $\beta
_j$s in a model for $Y$ given $\mathbf X$ were unrelated to the $p_j$s. In
that case $\mathit{EV}_j$ and $\mathrm{nc}_j$ tend to be smaller for rarer variants and a
smaller weight is preferred. However, if larger $|\beta_j|$s are more
likely to be found among rarer variants, then $W_H$ could be more
powerful than $W_C$. Simulations in Section~\ref{sec4} confirm this.

Powers of the linear statistics depend on the effect directions and on
the weights.
% The optimal weight is $w_j =\beta_j/\sigma^2$, but it is unattainable
%in practice without (correct) prior information, so
The effect of using weights inversely proportional to $p_j$ [e.g., $W_{Lp}=\mathbf w' S$ with $w_j=1/\sqrt{p_j(1-p_j)}$] is unclear, because
$\mathrm{nc}_{Lp}$ in (\ref{eq:ncLp})
is not necessarily bigger than $\mathrm{nc}_{L1}$ in (\ref{eq:ncL1}) for
$W_{L1}$ with equal weights,
even if rarer variants tend to have bigger genetic effects in terms of
larger $|\beta|$ values. We provide numerical results on the power of
$W_{L1}$, $W_{Lp}$, $W_{C}$ and $W_{H}$ under various conditions in
Section~\ref{sec4} for studies of both quantitative and binary traits.

%s3.3 #&#
\subsection{Additional Theoretical Results with More General Settings}\label{sec3.3}

Here we investigate the effects of dependency between genotypes.
Due to genetic linkage, rate of recombination, genetic selection and
other factors, genotypes of SNPs from the same chromosomal region may
not be independent of each other at the population level, that is,
$P(X_{ij} X_{ij'}) \neq P(X_{ij})P(X_{ij'})$. This phenomenon is also
known as linkage disequilibrium [e.g., \citet{Reich}]. Similar to the
previous section, we discuss results based on linear normal model (\ref
{eq:model}) and score statistic $\mathbf S=(S_1,\ldots,S_J)'$ in (\ref
{eq:Sj2}). This statistic can be rewritten in vector form as
%
%e3.11 #&#
\begin{equation}
\mathbf S = X'_c \mathbf Y,
\end{equation}
where $ X_c$ has $(i,j)$ entry $X_{ij} - \overline{X}_j$ (where $\overline
{X}_j = \sum_{i=1}^{n}X_{ij}/n $). Due to normality of $Y$, the
distribution of $\mathbf S$ given genotypes $\mathbf X$ is multivariate normal,
%
%e3.12 #&#
\begin{equation}
\mathbf S \sim N(\bolds\mu,\Sigma),
\end{equation}
where $\bolds\mu= E(\mathbf S) = X' X_c\bolds\beta= X'_c X_c\bolds\beta$ and
$\operatorname{Var}(\mathbf S)=\Sigma= \sigma^2 X'_c X_c$. We denote $n\hat{\Sigma}_X =
X'_c X_c$, an estimate of the covariance matrix of genotypes $\mathbf X$
and so $\bolds\mu= n\hat{\Sigma}_X\bolds\beta$ and $\Sigma= \sigma^2 n\hat
{\Sigma}_X$. Under mutually independent genotypes, matrix $\Sigma_X$ is
approximately diagonal, $n\hat{\Sigma}_X=\operatorname{diag}\{
m_1(1-m_1/n),\ldots,m_J(1-m_J/n)\}$, and we provided insights on the
effect of the choice of linear and quadratic statistics for this
covariance structure in Section~\ref{sec:normalsub}. Here we give additional results
for the general covariance structure. Similar to the previous sections,
$m_j$ and $m_{lj}=\sum_{i=1}^{n}X_{il}X_{ij}$ are treated as fixed
values, and for simplicity we consider the case where $m_j$ is equal to
its expected value $np_j$ and $m_{lj}$ is equal to its expected value
$np_{lj}$, where $p_{lj}=P(X_{il}=1,X_{ij}=1)$.

Similar to the previous section, we consider settings according to the
variation of $Y$ explained by the set of SNPs. Under model (\ref
{eq:model}) and covariance structure $\Sigma_X$, the total phenotypic
variation explained by the $J$ SNPs is
%
%e3.13 #&#
\begin{eqnarray}\label{eq:EVc}
\mathit{EV} &=& \frac{\operatorname{Var}(E[Y|\mathbf X])}{\operatorname{Var}(Y)} = \frac{\bolds\beta' \Sigma_X \bm
\beta}{\bolds\beta' \Sigma_X \bolds\beta+\sigma^2 }
\nonumber
\\[-8pt]
\\[-8pt]
\nonumber
&\approx&\frac{\bolds\beta
' \Sigma_X \bolds\beta}{\sigma^2 }
\end{eqnarray}
when explained variation is small. One should note that when genotypes
are not mutually independent, the total explained variation by $J$ SNPs
is not approximately equal to the sum of the individual explained
variations as in (\ref{eq:EV}).

Again we consider the two linear statistics $W_{L1} = \mathbf w' \mathbf S$
with $w_j=1$, $W_{Lp}$ with $w_j=1/\sqrt{p_j(1-p_j)}$ and two quadratic
statistic $W_Q = \mathbf S' A \mathbf S$: $W_C$ with $A=I$ (C-alpha) and
Hotelling $W_H$ with $A=\Sigma$. We note again that we are considering
the situation where the values of $m_j$ and $m_{lj}$ are equal to their
expected values $np_j$ and $np_{lj}$, respectively, thus, $\hat
{p}_j=p_j$ and $\hat{\Sigma}_X=\Sigma_X$. Let $\mathbf U\Lambda\mathbf U'$ be
the eigendecomposition of matrix $\Sigma_X$, where $\Lambda=\operatorname{diag}\{
\lambda_1,\ldots,\lambda_J\}$ consists of the eigenvalues of $\Sigma_X$
and $\mathbf U = \{\mathbf u_1,\ldots,\mathbf u_J \}$ is an orthogonal matrix
constructed from corresponding eigenvectors $\mathbf u_1,\ldots,\mathbf u_J$.
Based on the derivations in Section~\ref{sec:2.2}, the following
distributional results hold:
\begin{longlist}[(iii)]
\item[(i)] $W^2_{L1}/(\sigma^2\mathbf1'\Sigma_X\mathbf1)\sim\chi^2_{1,\mathrm{nc}}$,
with noncentrality parameter $\mathrm{nc} = n\frac{(\mathbf1' \Sigma_X\bolds\beta
)^2}{\sigma^2\mathbf1'\Sigma_X\mathbf1}$.
\item[(ii)] $W^2_{Lp}/(\sigma^2\mathbf w'\Sigma_X\mathbf w)\sim\chi^2_{1,\mathrm{nc}}$,
with noncentrality parameter $\mathrm{nc} =n \frac{(\mathbf w' \Sigma_X\bolds\beta
)^2}{\sigma^2\mathbf w'\Sigma_X\mathbf w}$ and $\mathbf w=(1/\sqrt {p_1(1-p_1)},\break  \ldots,1/\sqrt{p_J(1-p_J)})'$.
\item[(iii)] $W_C \sim\sum_{j=1}^{J}\lambda_j\chi^2_{1,\mathrm{nc}_j}$, with
$\mathrm{nc}_j=n\lambda_j(\mathbf u'_j \bolds\beta)^2/ \sigma^2$.
\item[(iv)] $W_H \sim\chi^2_{\operatorname{rank}(\Sigma_X),\mathrm{nc}}= \sum_{j=1}^{J}I(\lambda_j>0)\chi^2_{1,\mathrm{nc}_j}$, with $\mathrm{nc}_j=n\lambda_j(\mathbf
u'_j \bolds\beta)^2$ and $\mathrm{nc}=\sum_{j=1}^{J}n\lambda_j(\mathbf u'_j\bolds\beta
)^2/\break \sigma^2 = n\bolds\beta' \Sigma_X \bolds\beta/ \sigma^2 \approx n \mathit{EV}$.
\end{longlist}

The power of the Hotelling statistic $W_H$ again depends solely on
(approximate) explained variation by the $J$ SNPs and $\operatorname{rank}(\Sigma_X) =
\sum_{j=1}^{J}I(\lambda_j>0)$. If two different sets of $J$ SNPs
explain the same total phenotypic variation, then the power for $W_H$
is the same for those two sets regardless of the correlation structure
between SNPs, provided the corresponding $\Sigma_X$s have the same
rank. This also implies that when two sets of $J$ SNPs explain the same
total phenotypic variation, the Hotelling statistic is more powerful
for the set of SNPs where $\Sigma_X$ has lower rank. A second
conclusion is that power of the other three statistics depends on the
covariance structure of the SNPs, $\hat{\Sigma}_X$, and their effects
$\bolds\beta$. In fact, when two sets of $J$ SNPs explain the same total
phenotypic variation and one of the sets consists of mutually
independent SNPs, the power of these three tests for the set of
independent SNPs is not necessary larger than the power for another set
of SNPs with a different covariance structure. This is confirmed by our
empirical evaluations presented in supplementary materials [\citet{Supp}].
\section{Numerical Power Comparisons}\label{sec4}
We conducted extensive and novel simulation studies to examine the
finite sample performance of linear and quadratic statistics. Since
there is little background information suggesting what genetic
scenarios are most plausible, we generated data from over 10,000
different genetic models that involve varying proportions of
protective, deleterious and neutral variants, variant frequencies,
effect sizes, and relationships between variant frequencies and effect
sizes. Careful analysis of the results provides considerable insight
into the performance of different statistics. The statistics considered
here are the two linear statistics, $W_{L1}=\mathbf1' \mathbf S$, $W_{Lp}=\mathbf
w' \mathbf S$, where $w_j=1/\sqrt{p_j(1-p_j)}$, and two quadratic
statistics $W_C= \mathbf S' I \mathbf S$ and $W_H=\mathbf S' \Sigma^{-1} \mathbf S$, as
discussed in Section~\ref{sec:normalsub} and Table~\ref{tab:new}. Estimation of the
$p_j$ is discussed in Sections~\ref{sec4.1} and \ref{sec4.2} below.

%t2 #&#
\begin{table*}
\caption{Parameters and parameter values of simulated
models for studies of quantitative or binary traits. Scenario S1
(MAF-effect independent) assumes MAFs and effect sizes are mutually
independent. Scenario S2 (MAF-effect dependent) assumes that variants
with smaller MAFs tend to have bigger effect sizes}
\label{tab:1}
\begin{tabular*}{\textwidth}{@{\extracolsep{\fill}}lcp{185pt}@{}}
\hline
&\textbf{Parameters}& \multicolumn{1}{c@{}}{\textbf{Parameter values}}\\
\hline
$n$ &Sample size ($n_{\mathrm{case}}=n_{\mathrm{control}}=n/2$ for binary traits) & 500,
1000 or 2000 \\
$J$& Total number of SNPs & $\operatorname{Unif} \{10, 20, 30, 40, 50\}$ \\
$p_C$ & Proportion of the causal SNPs& $\operatorname{Unif} (0.1, 1)$ \\
$J_C$ & Number of the causal SNPs, an integer closest to $J \cdot p_C$&
\\
$p_D$ & Proportion of the deleterious SNPs among the causal ones & $\operatorname{Unif}
(0.75, 1)$\\
$J_D$ &Number of the deleterious SNPs, an integer closest to $J_C \cdot
p_D$ & \\
$p_P$ & Proportion of the protective SNPs among the casual ones,
$1-p_D$ \\
$J_P$ & Number of the protective SNPs, $J_C-J_D$ & \\
$p_N$ & Proportion of the neutral SNPs, $1-p_C$ &\\
$J_N$ & Number of the neutral SNPs, $J-J_D-J_P$ &\\[6pt]
\multicolumn{3}{c}{\textit{Quantitative traits under scenario \textup{S1}
\textup{(}MAF-effect independent\textup{)}; \textup{10,000} independently simulated models}}\\
$p_j$ & Approximately twice the MAF of SNP $j$ & $\operatorname{Unif} (0.005, 0.02)$\\
$\beta_j$ & Regression coefficient in (\ref{eq:model}) of SNP $j$ &\\
& for neutral SNPs & 0\\
& for causal SNPs & $\operatorname{Unif} (0.45, 0.5)$ or
$\operatorname{Unif} (-0.5, -0.45)$\\
& & (The resulting $\mathit{EV}_js$ in the range 0.001 to 0.0049) \\ [6pt]
%& for deleterious SNPs & \operatorname{Unif} (0.045, 0.5) \\
%& for protective SNPs & \operatorname{Unif} (-0.05, -0.045)\\ \hline
\multicolumn{3}{c}{\textit{Quantitative traits under scenario \textup{S2}
\textup{(}MAF-effect dependent\textup{)}; \textup{10,000} independently simulated models}}\\
%$\sigma^2$& total phenotypic variance (does not affect power; see text
%for details)&1 \\
$\mathit{EV}_j$ & The variance explained by SNP $j$ ($\mathit{EV}_j=\beta_j^2
p_j(1-p_j)$)& \\
& for neutral SNPs & 0\\
& for causal SNPs& $\operatorname{Unif} (0.001, 0.0025)$ \\ [6pt]
\multicolumn{3}{c}{\textit{Binary traits under scenario \textup{S1} \textup{(}MAF-effect
independent\textup{)}; \textup{500} independently simulated models}}\\
$p_j$ & Approximately twice the MAF of SNP $j$ & $\operatorname{Unif} (0.005, 0.02)$\\[2pt]
$e^\beta_j$ & OR of SNP $j$ &\\
& for neutral SNPs & 1\\
& for causal SNPs & $\operatorname{Unif} (2, 4)$ or $\operatorname{Unif} (1/2, 1/4)$ \\[6pt]
\multicolumn{3}{c}{\textit{Binary traits under scenario \textup{S2} \textup{(}MAF-effect
dependent\textup{)}; \textup{500} independently simulated models}}\\
$p_j$ & Approximately twice the MAF of SNP $j$ & $\operatorname{Unif} (0.005, 0.02)$\\[2pt]
$e^\beta_j$ & OR of SNP $j$ &\\
& for neutral SNPs & 1\\
& for causal SNPs & $C/\sqrt{p_j(1-p_j)}, C=4\sqrt{0.005(1-0.005)}$ \\
& & (The resulting ORs in the range 2 (or $1/2$) to 4 (or $1/4$) \\
\hline
\end{tabular*}
%
%models.}
\end{table*}
%

%The `` MAF-effect independent'' assumption assumes that MAFs and effect
%sizes are mutually independent and ``MAF-effect dependent'' assumes
%that variants with smaller MAFs tend to have bigger effect sizes.
We studied both quantitative and binary traits. Table~\ref{tab:1}
describes the simulation models considered. For each type of trait, we
considered two types of scenarios, S1 (``MAF-effect independent'')
assumes that $|\beta_j|$ (the size of the genetic effect) of a causal
SNP $j$ is unrelated to $p_j$ (approximately twice the MAF), and S2
(``MAF-effect dependent'') assumes that
$|\beta_j|$ is inversely related to $p_j$. For normally distributed
quantitative traits, the MAF-effect dependent models were simulated by
directly specifying the phenotypic variance explained by SNP $j$, $\mathit{EV}_j
= (\beta_j \sqrt{p_j (1-p_j)})^2/\sigma^2$, and without loss of
generality we take $\sigma^2=1$.
We did not restrict all causal variants to have the same direction of
effect, but assumed that the majority of the causal variants have the
same direction with $p_D=J_D/J_C$ ranging from 75\% to 100\%, a
reasonable assumption based on what has been reported in the
literature. (We also simulated models where $p_C$ ranges from 50\% to
75\%; the linear statistics performed poorly and were dominated by the
quadratic statistics, as one would expect.)
Here we assume that the genotypes of different SNPs are mutually
independent, but Section~\ref{sec5} considers possibly nonindependent genotypes
obtained from sequence data of the 1000 Genomes Project [\citet
{1000GenomesProjectConsortium}]. We also conducted additional
simulation studies examining the effect of dependency between SNPs on
power, supporting conclusions made in Section~\ref{sec3.3} above.

%s4.1 #&#
\subsection{Quantitative Traits}\label{sec4.1}
We first considered the normal linear model in (\ref{eq:model}) for
which results in Section~\ref{sec:normalsub} give the power of the different statistics.
Results presentation and discussion focus on $n=1000$ and type 1 error
$\alpha=10^{-4}$.
(Other $n$ and $\alpha$ values were also considered, but results are
qualitatively similar across tests.) The choice of $\alpha=10^{-4}$ is
to reflect the fact that testing would typically be conducted for
multiple genetic regions. Table~\ref{tab:1} shows the combination of
factors and indicates how data from 10,000 different models were generated.

%f1 #&#
\begin{figure*}

\includegraphics{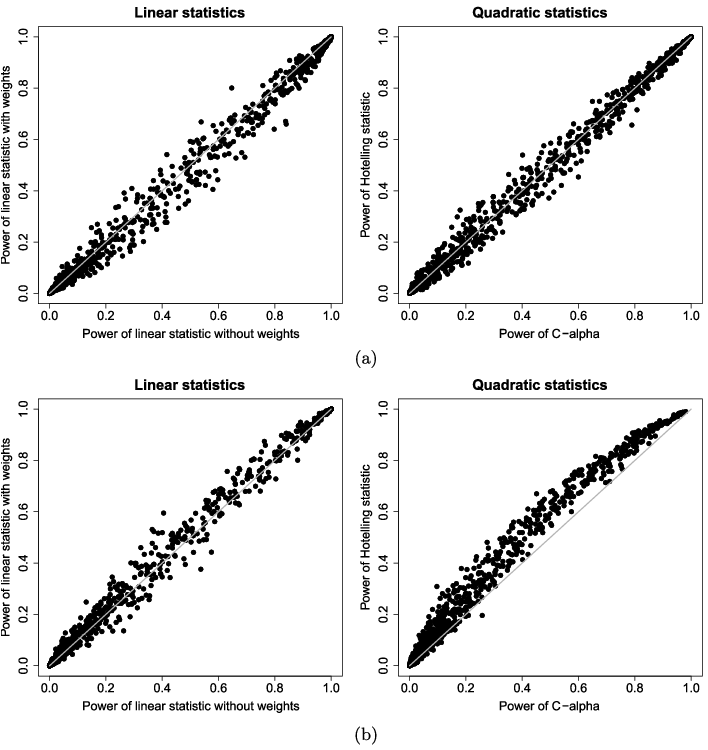}

\caption{Within-class power comparison of the four statistics
for 1000 independently generated models for studies of {QUANTITATIVE}
traits under \textup{(a)} scenario S1 (MAF-effect independent) and \textup{(b)}
scenario
S2 (MAF-effect dependent) as described in Table \protect\ref{tab:1}. The four
statistics are
the two linear statistics $W_L=(w_1,\ldots, w_J)'\mathbf S$ in (\protect\ref
{eq:WL}): ``without weights'' $W_{L1}$ where $w_j\equiv1$ and ``with
weights'' $W_{Lp}$ where $w_j=1/\sqrt{p_j(1-p_j)}$, and
two quadratic statistics $W_Q=\mathbf S' A \mathbf S$ in (\protect\ref{eq:WQ}): the
C-alpha statistic $W_C$ where $A=I$ and the Hotelling statistic $W_H$
where $A=\Sigma^{-1}_S$. Sample size $n=1000$ and type 1 error $\alpha
=10^{-4}$. The set of 1000 models presented here is a random subset of
all the 10,000 models independently generated.}\label{fig: 1}  %\caption{}\label{}
\end{figure*}

For each of the 10,000 randomly generated genetic models we used
critical values according to the exact distributions in Section~\ref{sec:normalD} to
compute power. Specifically, for each model we considered a sample of
size $n=1000$ for which the $m_j$ equaled their expected values $np_j$.
Thus, $\hat{p}_j=p_j$ for each SNP and the $J$ by $J$ covariance matrix
$\Sigma$ in (\ref{eq: 4.4}) equals $\operatorname{diag}\{np_j(1-p_j)\sigma^2\}$ under
both the null ($\bolds\beta=0$) and alternative hypothesis represented by
the genetic model. Since $n$ is large, we ignored the effect of
estimating $\sigma^2$ (as in Section~\ref{sec:normalD}) and used the true value
$\sigma^2=1$; this has a negligible effect on power. The use of $\hat
{p}_j = p_j$ deserves discussion, since in practice the value $\hat
{p}_j$ will vary from sample to sample. However, they are functions
only of the covariates $X_{ij}$ and so no adjustments to the
distribution in Section~\ref{sec:normalD} are needed. However, the power provided by
using (\ref{eq:WL2}) or (\ref{eq:WQ2}) with the $p_j$ estimated with
$\hat{p}_j$ are conditional, that is, they apply to samples with the
described set of values $m_j$. Unconditional power is also of interest;
this reflects sampling variation in the $m_j$ (and $\hat{p}_j$).
Unconditional power is calculated (or estimated) by averaging
conditional powers for the case where $m_j = np_j$ in this section. In
the supplementary materials [\citet{Supp}] we provide some
unconditional power values. We find that differences with the
conditional powers are small (see Figures S6 and S7).

%f2 #&#
\begin{figure*}

\includegraphics{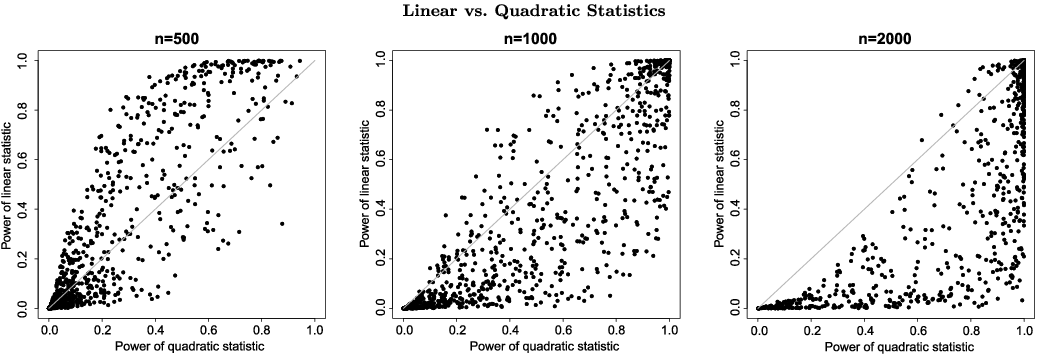}

\caption{Between-class power comparison of the linear statistic
$W_{L1}$ vs.  the quadratic Hotelling statistic $W_H$ for studies of
{QUANTITATIVE} traits under scenario S1 (MAF-effect independent).
Other details see Figure \protect\ref{fig: 1}.}\label{fig:2} % \caption{}\label{}
\end{figure*}
%{~~~~~~~~ \textbf{Linear vs. Quadratic Statistics} \\}

%
%t3 #&#
\begin{table*}[t!]
\caption{Breakdown of the power of the linear
statistic $W_{L1}$ and the quadratic Hotelling statistic $W_H$ under
scenario S1 (MAF-effect independent). Proportions of the 1000 models
in Figure \protect\ref{fig:2} that have power in the specified ranges. For
other details see Figures \protect\ref{fig: 1} and \protect\ref{fig:2} legends}
\label{tab:2}%
\begin{tabular*}{\textwidth}{@{\extracolsep{\fill}}lccccc@{}}
\hline
 & \multicolumn{5}{c@{}}{\textbf{Power range}}\\
\cline{2-6}
\multicolumn{1}{@{}l}{\textbf{Sample size}} & \textbf{0--20\%} & \textbf{20--40\%} &
 \textbf{40--60\%} & \textbf{60--80\% }& \textbf{80--100\%} \\
\hline
\multicolumn{6}{c}{Proportion of the
models in power range; $W_{L1}$}\\
$n=500$ & 0.66 & 0.11 & 0.06 & 0.06 & 0.11 \\
$n=1000$ & 0.46 & 0.11 & 0.08 & 0.07 & 0.28 \\
$n=2000$ & 0.30 & 0.08 & 0.06 & 0.07 & 0.49 \\[3pt]
\multicolumn{6}{c}{Proportion of the
models in power range; $W_{H}$}\\
$n=500$ & 0.68 & 0.14 & 0.09 & 0.07 & 0.02 \\
$n=1000$ & 0.32 & 0.13 & 0.10 & 0.10 & 0.35 \\
$n=2000$ & 0.10 & 0.07 & 0.06 & 0.07 & 0.70 \\
\hline
\end{tabular*}
\end{table*}
%

%f3 #&#
\begin{figure*}

\includegraphics{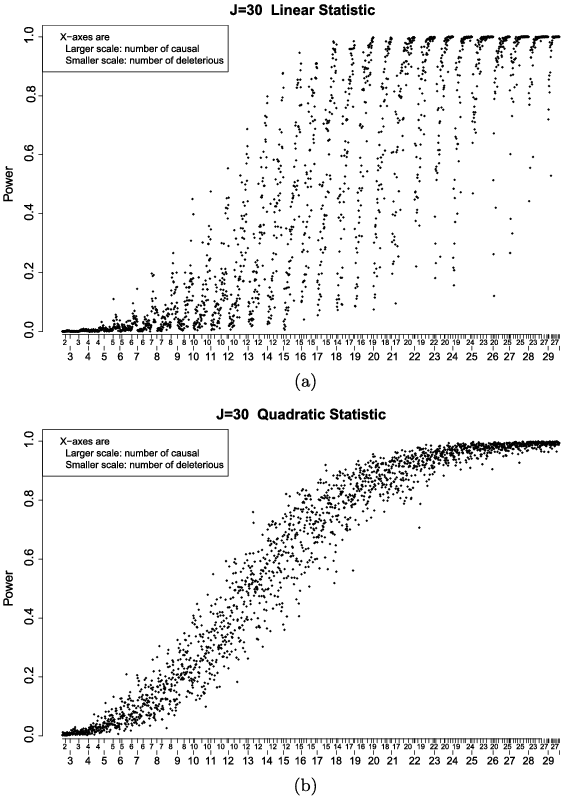}

\caption{Individual power of \textup{(a)} the linear statistic $W_{L1}$
and \textup{(b)} the quadratic Hotelling statistic $W_H$ for studies of
{QUANTITATIVE} traits under scenario S1 (MAF-effect independent)
for models with $J=30$ total number of rare variants. The large scale
of the $X$-axis shows the number of causal variants in the range of
$J_C=J \cdot p_c=30 \cdot10\%=3$ to $J_C=30 \cdot100\%=30$. The small
scale of the $X$-axis shows the number of deleterious variants in the
range of $J_D=J_C \cdot p_D=J_C \cdot75\%$ to $J_D=J_C\cdot100\%$,
depending on the actual number of causal variants in a model. The 2005
models shown here are the models with $J=30$ among the 10,000 models
generated as described in Table \protect\ref{tab:1}. Sample size $n=1000$ and
type 1 error $\alpha=10^{-4}$.}
\label{fig: 3} % \caption{}\label{}
\end{figure*}

For visual display, Figure~\ref{fig: 1} shows the within-class power
comparisons (linear $W_{Lp}$ vs. linear $W_{L1}$, and quadratic $W_H$
vs. quadratic $W_C$) of the four tests for 1000 models randomly
selected from the 10,000 independently generated models. In view of the
wide variations in model parameters, powers of the tests vary widely
across the 1000 models. For each model, powers of the two linear
statistics are similar and likewise for powers of the two quadratic
statistics. Moreover, under scenario S1 [Figure~\ref{fig: 1}(a)] neither
statistic within each class dominates the other across the 1000 models.
However, under scenario S2 [Figure~\ref{fig: 1}(b)], the Hotelling
statistic performs better than the C-alpha statistic for almost all
models, as our earlier comments in Section~\ref{sec:normalsub} suggest. In this case,
we also see that the linear statistic using weights inversely
proportional to MAFs does not always lead to a better power even when
the assumption that rarer variants have bigger effects is in fact true
here [Figure~\ref{fig: 1}(b)].

We also considered simulations with sample sizes $n=500$ and 2000, to
see the effect on the linear versus quadratic statistic comparison. For
simplicity we show plots for $W_{L1}$ and $W_H$; plots for $W_{Lp}$ and
$W_C$ are very similar. Figure~\ref{fig:2} and Table~\ref{tab:2} show
that which type of statistic is better depends on the sample size and
the model parameters. When $n=500$, both the linear and quadratic
statistics have low power (more than 65\% of the 1000 models have power
$<$20\%; Table~\ref{tab:2}). In that case, good power (80\%) is
achieved only for those models with high proportions of causal SNPs
(among which the proportion of deleterious SNPs is at least 75\% by
study design); the linear statistic is better than the quadratic
statistic. However, as $n$ increases, the quadratic statistic displays
good power across many models and by $n=2000$ dominates the linear
statistic for most of the models. Similar conclusions can be made based
on results from the models simulated under scenario S2 (see
supplementary materials Figure S1 [\citet{Supp}]).

To better understand the impact of the various model parameters on
different statistics, Figure~\ref{fig: 3} presents power from a
different perspective showing the individual power of the linear
statistic $W_{L1}$ [Figure~\ref{fig: 3}(a)] and the quadratic statistic
$W_{H}$ [Figure~\ref{fig: 3}(b)] as a function
of the number of causal variants $J_C$ (large scale of the $X$-axis) and
the number of
deleterious variants $J_D$ (small scale of the $X$-axis), when the total
number of rare
variants is $J = 30$ under the scenario S1. Results for scenario S2 are
in supplementary materials Figure S2; results for $J =$ 10, 20, 40 and
50 are qualitatively similar and not shown. It is clear that the power
of both tests depends highly on the percentage of causal SNPs in the group
of SNPs investigated. For example, among the 10,000 models giving
power of 50\% or greater, the average proportion of causal SNPs ($p_C$)
is 81\% ($\mathrm{SE} = 13$\%
and $\min = 42$\%) for the linear test and 81\% ($\mathrm{SE} = 12$\% and $\min = 50$\%)
for the quadratic
test. The powers for the quadratic statistics vary much less than those
for the linear statistics; this is due to the latter's need for both
$p_C$ and $p_D$ (the proportion of deleterious SNPs among the causal
ones) being close to 1 in order to achieve high power.\looseness=1

To examine the effect of correlation between SNPs on power, we
conducted additional simulation studies. Briefly, we considered two
types of correlation scenarios (D1: correlation among casual variants
and D2: correlation between causal and neutral variants) and compared
power of the four tests ($W_{L_1}, W_{L_p}, W_C, W_H$) to the
independence case, under two different assumptions of the corresponding
genetic effects (E1: total explained variation by all causal variants
is fixed and E2: the regression coefficient $\beta_j$s are fixed).
Under E1, neither correlation structure affects power of $W_H$;
however, D1 increases power of the other three tests while D2 can
increase or decrease power. Under E2, D1 increases power of all four
tests; D2 once again can increase or decrease power. Details of the
simulation study design and results (Figures S8--S11) are in the
supplementary material [\citet{Supp}].

%f4 #&#
\begin{figure*}

\includegraphics{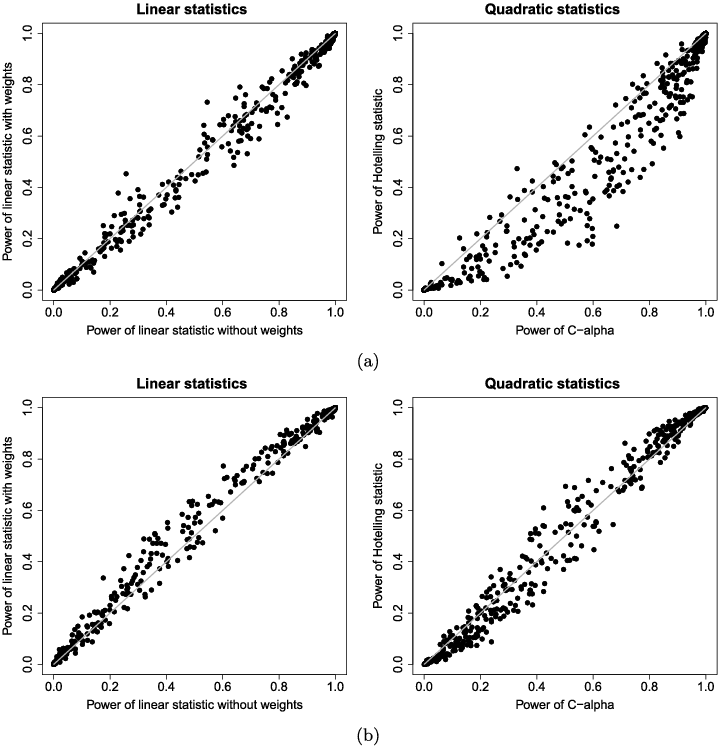}

\caption{Within-class power comparison of the four statistics
for 500 independently generated models for studies of BINARY traits
under \textup{(a)} scenario S1 (MAF-effect independent) and \textup{(b)} scenario S2
(MAF-effect dependent) as described in Table \protect\ref{tab:1}. The four
statistics are
the two linear statistics $W_L=(w_1,\ldots, w_J)'\mathbf S$ in (\protect\ref
{eq:WL}): ``without weights'' $W_{L1}$ where $w_j\equiv1$ and ``with
weights'' $W_{Lp}$ where $w_j=1/\sqrt{p_j(1-p_j)}$, and
two quadratic statistics $W_Q=\mathbf S' A \mathbf S$ in (\protect\ref{eq:WQ}): the
C-alpha statistic $W_C$ where $A=I$ and the Hotelling statistic $W_H$
where $A=\Sigma^{-1}_S$. Sample size $n=1000$ and type 1 error $\alpha
=10^{-4}$.}\label{fig: 4}
\end{figure*}
%s4.2 #&#
\subsection{Binary Traits}\label{sec4.2}

Here, we provide detailed numerical results for case--control studies
involving a binary trait $Y$, where a normal approximation for $\mathbf S$
might not be adequate. As in Section~\ref{sec4.1}, we examine the performance of
$W_{L1}$, $W_{Lp}$, $W_C$ and $W_H$. We assume that the distribution of
$Y_i$ given $\mathbf X_i=(X_{i1},\ldots,X_{iJ})'$ is Bernoulli with
%
%e4.1 #&#
\begin{equation}\quad
\operatorname{Prob}(Y_i=1|\mathbf X_i) = \frac{\exp(\beta_0+\sum\beta
_jX_{ij})}{1+\exp(\beta_0+\sum\beta_jX_{ij})},
\label{eq:modelcc}
\end{equation}
and that the $X_{ij}$s in the population are mutually independent
Bernoulli variables with $P(X_{ij}=1) =p_j$ for $j=1,\ldots,J$.
We first used asymptotic distributions for the linear and quadratic
statistics provided in Section~\ref{sec2.3} to obtain $p$-values, and we evaluated
type I error rate and obtained empirical critical values for each of
the four tests (supplementary materials Table S1). In this case the
test statistics are based on (2.3) with the covariance matrix given by
(A.3) in the supplementary materials [\citet{Supp}]. Unlike the
quantitative traits above, the SNP genotypes $X_{ij}$ here vary from
sample to sample and thus so do the values $\hat{p}_j$ $(j =1,\dots
,J)$. Supplementary Table S1 shows that normal approximations are
satisfactory for the linear statistics but chi-square approximations
for the quadratic statistic produce $p$-values (and thus critical values)
that are much too conservative.
We conducted simulations to assess power under different scenarios,
using empirical critical values for the quadratic statistics. The
simulation of case--control data is discussed in the  supplemetary
materials [\citet{Supp}]. Given the amount of computation required, we
considered 500 models randomly generated under each of the two
MAF-effect scenarios described in Table~\ref{tab:1}.

Results in Figure~\ref{fig: 4} are slightly different from those in
Figure~\ref{fig: 1} for quantitative traits. Under scenario S1
[Figure~\ref{fig: 4}(a), left panel], neither of the two linear statistics
dominates the other, which is similar to the case for quantitative
traits [Figure~\ref{fig: 1}(a), left panel]. Between the two quadratic
statistics [Figure~\ref{fig: 4}(a), right panel], $W_C$ is more powerful
than $W_H$; this is consistent with the findings of \citet{BasuandPan}
discussed in Section~\ref{sec:normalsub}. However,
the systematic power difference between $W_C$ and $W_H$ is absent
under scenario S2 [Figure~\ref{fig: 4}(b), right panel]. This supplements
the picture provided by \citet{BasuandPan}, who did not consider cases
where genetic effects are inversely proportional to MAFs, and it
supports our earlier comment that the relative performance of $W_C$ and
$W_H$ depends on the relationship between SNP effects and
MAFs.\looseness=1

Under the MAF-effect dependent assumption, the linear statistic
$W_{Lp}$ appears to be consistently better than $W_{L1}$ across the 500
models [Figure~\ref{fig: 4}(b), left panel]. However, we emphasize that
the apparent better power for $W_{Lp}$ is mainly driven by the use of
true variant frequency $p_j$ values in the weight specification,
$w_j=1/\sqrt{p_j(1-p_j)}$. These would be unavailable to us in a real
situation. In practice, how to estimate $p_j$ can have major impacts on
the validity of the test as well as on power. Some authors have
suggested using the control sample only [e.g.,  \citet
{MadsenandBrowning}], but it is not clear if the standard
permutation-based approach for $p$-value estimation as used here is still
valid. An additional concern for this approach is the possibility of a
deleterious effect. In that case, which subsample is the proper
``control'' sample is not clear. If both cases and controls were used to
estimate $p_j$, $\hat p_j$ would tend to be bigger than $p_j$ for a
causal SNP $j$ because of the oversampling of cases, while $\hat
p_{j'}$ is expected to be $p_{j'}$ for a neutral SNP~$j'$.
Consequently, using
$w_j=1/\sqrt{\hat p_j(1-\hat p_j)}$ downweights a causal SNP compared
to a neutral one with the same frequency, resulting in loss of power.
This is clear from the results shown in supplementary materials Figure~S3
for both the MAF-effect independent and dependent scenarios. The
practical use of weights, particularly for linear statistics,
therefore, must be carefully considered in the case--control setting.

\begin{figure*}

\includegraphics{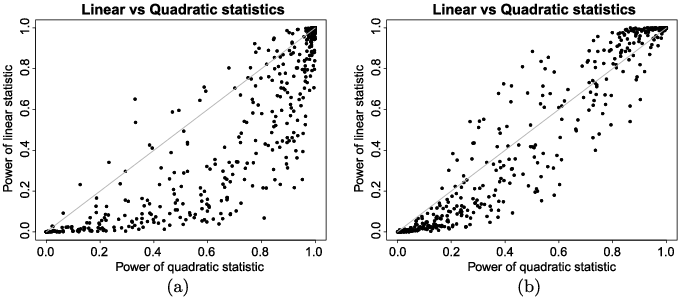}

\caption{Between-class power comparison of the two statistics
for 500 independently generated models for studies of BINARY traits
under \textup{(a)} scenario S1 (MAF-effect independent) and \textup{(b)} scenario S2
(MAF-effect dependent) as described in Table \protect\ref{tab:1}.
The linear statistic is $W_{L1}$ and the quadratic statistic is C-alpha
statistic $W_C$. For other details see Figure \protect\ref{fig: 4} legends.}
\label{fig: 5}  %\caption{}\label{}
\end{figure*}

Figure~\ref{fig: 5} compares the power of $W_{L1}$ and $W_{C}$ across
the 500 models.
Under scenario S1 [Figure~\ref{fig: 5}(a)],
the quadratic statistic has better power than the linear statistic for
the majority of the models.
Under scenario S2 [Figure~\ref{fig: 5}(b)],
among the models with power less than $50\%$,
the quadratic statistic has better power, but
among the models with higher power, the linear statistic is more often
better.\looseness=1
%This is because, given the sample size of $n_0=n_1=500$, high power is
%achieved mainly when $p_C$ and $p_D$ are both close to 1, in which
%case linear statistics do well. Patterns change with larger sample or
%effect sizes as in Figure 2.

%f5 #&#

%s5 #&#
\section{Application to the GAW17 Data}\label{sec5}
The numerical studies in the previous section focused on mutually
independent SNPs,
although the tests themselves do not require this [see supplementary materials
(\citet{Supp}] for additional simulation studies on dependent SNPs). To consider
settings where this might not be so along with real sequence data, we
examined real
human sequence data [\citet{1000GenomesProjectConsortium}] that
were used to generate
the GAW17 phenotype data [\citet{Almasyetal}] introduced in Section~\ref{sec1}.

We consider here quantitative trait Q2 which is influenced by 72 SNPs
in 13 genes but not by other covariates; recall from Section~\ref{sec1} that
traits were simulated, so it is known which SNPs are causal. To \mbox{assess}
the performance of association statistics, we carried out ``pseudo
power'' comparisons by determining the $p$-values for each of four test
statistics, across each of the 13 genes, using the 200 replicate
samples available (same genotype data but different phenotype data,
independently simulated, based on the true genotype--phenotype
association model).

We used data from the $n=321$ unrelated Asian subjects (Han Chinese,
Denver Chinese and Japanese) and excluded SNPs that had MAF $>$5\% or
were monomorphic within the Asian sample. Gene \textit{VNN1} had no
causal rare variant but it was kept in the analysis to serve as a
negative control. The threshold MAF $\leq$5\% does not reduce the number
of causal SNPs much (70 of the 72 causal SNPs have MAF $\leq$5\%), but
it reduces the number of neutral SNPs in a gene and therefore increases power.

For each of the 200 replicates, we calculated permutation-based
$p$-values for the four statistics, $W_{L1}$, $W_{Lp}$, $W_{C}$ and
$W_{H}$ (see Table~\ref{tab:new}). We estimated power for $\alpha=0.05$
by the proportion of the 200 replicates for which the empirical
$p$-values were $\leq0.05$ for each test. For each sample, gene and
statistic combination, the $p$-value for the null hypothesis of no
association was obtained from the permutation distribution by randomly
generating 10,000 permutations of each replicate sample.

The choice of the liberal type 1 error $\alpha=0.05$ was based on the
low power of detecting genetic effects of sizes represented by the
simulation models, with a sample of 321 people. Table~\ref{tab:3}
summarizes the rare variants for the 13 genes and gives the empirical
power for each statistic. Only the first group of 9 genes have maximum
power above 10\%.

%t4 #&#
\begin{table*}
\tabcolsep=5pt
\caption{Power of the four test statistics applied to
the GAW17 sequence data provided by the 1000 Genomes Project.
The 13 genes presented here are all the causal genes for simulated
quantitative trait Q2.
\textit{VNN}1 does not have causal variants because one of the two
causal variants has MAF 26\% and the other is not polymorphic within
the Asian sample ($n=321$).
\textit{VNN}1 is kept in the analysis to serve as a negative control.
All causal variants were designed by GAW17 to have the same direction
of effects (minor alleles were associated with higher Q2 values). The
average genetic effect is the average of regression coefficient $\beta$
values of the causal variants used to simulate Q2 (effects are
independent of populations by the GAW17 design).
%Powers are estimated using all 200 replicates at the $\alpha=0.05$
%level for each of the four tests.
Genes are ordered according to the maximum power of the four tests
which is bolded.
Powers shown vary considerably due to inherent factors and estimation
based only on 200 replicates, and the 13 genes are separated into
different groups}\label{tab:3}
\begin{tabular*}{\textwidth}{@{\extracolsep{\fill}}lccccccc@{}}
\hline
& \textbf{SNP distribution}& \textbf{Ave. MAF of} & \textbf{Avg. effect of} &
\multicolumn{4}{c}{\textbf{Power}}\\
\cline{5-8}
\textbf{Gene}& $\bolds{J_C, J_N}$ & $\bolds{J_C, J_N}$ &
$\bolds{J_C}$ & \textbf{Linear} $\bolds{W_{Lp}}$ & \textbf{Linear}
$\bolds{W_{L1}}$& \textbf{Quadratic} $\bolds{W_C}$ &\textbf{Quadratic} $\bolds{W_H}$\\
\hline
\multicolumn{8}{c}{9 genes for which the maximum power is 10\% or
more}\\
$\mathit{SIRT}1$& 4, 7 &0.27\%, 0.22\%&0.71&\textbf{0.44}&0.40& 0.25& 0.39\\
$\mathit{BCHE}$& 5, 10 &0.22\%, 0.19\%&0.72&0.29&0.35& \textbf{0.43}& 0.39\\
$\mathit{PDGFD}$& 3, 6 &0.78\%, 0.65\%&0.74&0.29&0.43&\textbf{0.45} & 0.35\\
$\mathit{SREBF}1$& 4, 5 &0.39\%, 0.40\%&0.52&\textbf{0.49}&0.47& 0.18&0.28\\
$\mathit{GCKR}$& 1, 0& 1.21\%, $ $ NA&0.38&\textbf{0.25}&\textbf{0.25}&\textbf
{0.25}&\textbf{0.25}\\
$\mathit{RARB}$& 1, 5 &0.78\%, 0.90\%&0.64&0.06&0.03& 0.07& \textbf{0.14}\\
$\mathit{PLAT}$& 4, 7 &0.39\%, 0.49\%&0.68&\textbf{0.13}&\textbf{0.13}& 0.06&
\textbf{0.13}\\
$\mathit{VLDLR}$& 4, 6 &0.19\%, 1.64\%&0.75&\textbf{0.12}&0.08& 0.06& 0.09\\
$\mathit{VNN}3$& 2, 2 &0.16\%, 2.57\%&0.37&0.03&\textbf{0.10}& 0.06& 0.04\\[3pt]
\multicolumn{8}{c}{3 genes for which the maximum power is 10\% or
less}\\
$\mathit{INSIG}1$& 3, 1 &0.16\%, 3.42\%&0.20&0.06&0.06& 0.04& 0.03\\
$\mathit{LPL}$& 1, 4 &0.16\%, 0.23\% &0.73&0.02&0.03& 0.06& 0.05\\
$\mathit{VWF}$ &1, 3 &0.16\%, 1.90\%&0.34&0.02&0.01& 0.03& 0.01\\ [3pt]
\multicolumn{8}{c}{1 gene for which there is no polymorphic rare causal
variants in the Asian sample}\\ [3pt]
$\mathit{VNN}1$ & 0, 3& NA, $ $0.31\% & NA& 0.02&0.02& 0.04& 0.05\\
\hline
\end{tabular*}
\end{table*}

Results in Table~\ref{tab:3} are consistent with our previous
conclusions: (i) linear tests with and without weights based on MAF
vary in relative power but not substantially; (ii) quadratic statistics
$W_C$ and $W_H$ also have slightly variable relative power; (iii)
between-class performance is highly variable. As expected, linear
statistics outperform quadratic statistics if the proportion of causal
variants is not too low (e.g., genes $\mathit{SIRT}1$ and $\mathit{SREBF}1$), but the
pattern can be reversed if this is not the case, even when the effects
in this data are all in the same direction (e.g., $\mathit{BCHE}$ and $\mathit{RARB}$).

%s6 #&#
\section{Discussion and Recommendations}\label{sec6}

We have reviewed and studied tests of association between rare variants
and phenotypes within a unified framework which gives theoretical
insights about the performance of the methods (Table~\ref{tab:new}).
Tests can have greatly varying power depending on the total number of
rare variants, the numbers of deleterious, protective and neutral
variants, the effect directions and the relationship between the effect
sizes and the MAFs of causal variants. When substantial numbers of both
deleterious and protective SNPs are present, quadratic test statistics
are much better. They can also outperform linear statistics in settings
where causal SNPs are all deleterious (or all protective), but a
substantial fraction of the SNPs are not associated with the phenotype.
However, our results also indicate that power to detect moderate levels
of association is not high unless sample sizes are very large or a high
proportion of the chosen SNPs are causal. Sequencing errors and other
caveats concerning the data will further decrease power. Cases where
power is substantial for smaller studies are predominantly ones where
SNPs are almost all deleterious or all beneficial, and it is the linear
test statistics that achieve highest power. %Consequently, it is
%critical to obtain relevant biological information that can guide the
%selection of SNPs, their placement in groups, and weighting strategy
%for pooled association testing of rare variants (e.g.
Consequently, the definition of a chromosomal region and selection of
SNPs within the region are critical to statistical inference regardless
of the specific test used. In practice, a chromosomal region can be a
gene, coding region of a gene or other types of genetic unit (e.g., a
group of SNPs that are in moderate or strong linkage disequilibrium of
each other); selection of SNPs within a region can be also based on
relevant biological information since not all SNPs are equal {a
priori} (e.g., some SNPs are believed to be more important than others
based on functional genomic annotation). Different choices could lead
to different statistical power [e.g., \citet{Kingetal}, \citet{gaw18}].

Our work complements that of \citet{BasuandPan}, and a brief comparison
is useful. They found similar results to ours in simulation studies for
case--control scenarios, concerning the performance of linear
statistics. Among the quadratic statistics, they found that the
C-alpha/SSU type statistic $W_C = \mathbf S' I \mathbf S$ was generally the
best and superior to the Hotelling statistics $\mathbf S' \Sigma^{-1}\mathbf
S$. However, their simulation scenarios did not include cases where
larger causal effects are associated with SNPs having smaller MAFs. Our
numerical studies [scenario S2 under the MAF-effect dependent
assumption in Table~\ref{tab:1}; Figure~\ref{fig: 1}(b) for quantitative
traits and Figure~\ref{fig: 4}(b) for binary traits] and investigation of
GAW17 data (Table~\ref{tab:3}) indicate the importance of the
MAF-effect independent or nonindependent assumption on the choice of a
good test statistic.

As an approach to rare variant testing in the absence of strong prior
information, we support the recommendation of \citet{BasuandPan} to
perform tests using both linear and quadratic statistics. In \citet
{derkach} we investigated tests based on Fisher's method and the
minimum-$p$ method [e.g., \citet{Owen}] for combing $p$-values from linear
and quadratic statistics. Such tests were shown to be robust across the
wide range of models considered here, in the sense of achieving power
that is close to that of the better of a linear and quadratic statistic
in a given setting. Comparisons were also made with the recent SKAT-O
statistic of \citet{Leeetal}, which considers the minimum $p$-value
across a class of statistics. The overall conclusion is that the
Fisher's method outperforms the individual linear and quadratic tests
as well as the minimum $p$-value approach, when the majority of the
causal variants has the same direction of effect; however, the minimum
$p$-value is better if (approximately) half of the causal variants are
deleterious and the other half are protective.
%for ``sequence kernel association test - optimal''.

It is beyond our scope here, but an empirical assessment of test
statistics that involve covariate adjustment would be valuable. In
addition, accurate and computationally efficient methods of obtaining
$p$-values deserve attention. Parametric bootstrap simulation [e.g.,
\citet{LinandTang}] can be used when sampling of individuals is random,
but when it is trait-dependent matters are more complicated. In the
case--control simulation for binary traits, for example, the sampling is
\mbox{effectively} for $X_i$ and other covariates $v_i$ given $Y_i$. Methods
that avoid detailed modeling of the distribution of ($X_i, v_i$) are
desired. Empirical assessment is also difficult for family based
association studies when samples are correlated. We hope to report on
this in a future communication.

Finally, we reiterate our remarks made in Section~\ref{sec1} concerning the
potential effects of sequencing errors. A~realistic assessment of their
scope and impact is called for.

\section*{Acknowledgments}
The authors would like to thank the Genetic Analysis Workshop 17
(GAW17) committee and the 1000 Genomes Project for providing the GAW17
application data, and Dr. Andrew Paterson for insightful discussions.
This work was supported by
the Natural Sciences and Engineering Research Council of Canada (NSERC)
and the Canadian Institutes of Health Research (CIHR) grants to LS,
NSERC to JFL, the
Ontario Graduate Scholarship (OGS) and the CIHR Strategic Training for Advanced
Genetic Epidemiology (STAGE) fellowship to AD, University of Toronto.
\textit{Conflict of Interest}: None declared.

\begin{supplement}[id=suppA]
\stitle{Pooled Association Tests for Rare Genetic Variants: A Review
and Some New Results}
\slink[doi,text={10.\break 1214/13-STS456SUPP}]{10.1214/13-STS456SUPP} %[doi,] - jei reikia suskaldyti doi
\sdatatype{.pdf}
\sfilename{sts456\_supp.pdf}
\sdescription{The supplementary materials include derivation of the
permutation distribution of $\mathbf S$ for general traits, analytical
results and simulation details for study of binary traits, simulation
details for study of the effect of correlation between SNPs on power,
and an additional 1 table and 11 figures for the studies of type 1
error rates and power for both quantitative and binary traits, for both
MAF-effect independent and dependent scenarios, and for both
independent and dependent rare variants.}
\end{supplement}

%{\small
%}

%
%
%}
%
%%
%
%

%%
%$
%%
%%
%$
%
%%
%

% Is Hotelling reference correct? Is Ling and Tang discussed, in
%principle, correctly here?



% imsref loaded by akundreckaite, 2013-12-23 14:36:30
%


\begin{thebibliography}{37}
% Style name=ims, version=2.4, label_style=nameyear,
%sorting_style=complex, cfg=None, language=None.

%b1 ###
%b1 #&#
\bibitem[\protect\citeauthoryear{1000 Genomes Project Consortium}{2010}]{1000GenomesProjectConsortium}
%
\begin{bmisc}[pbm]
\borganization{1000 Genomes Project Consortium}
(\byear{2010}).
\bhowpublished{A map of human genome variation from population-scale sequencing.
\textit{Nature}
\textbf{467}
1061--1073.}
%pii={nature09534}, pmcid={3042601}, pmid={20981092}}
\end{bmisc}
%
\bptok{imsref}%
% NOT OUTPUTED:
% issn = 1476-4687
% number = 7319
\endbibitem

%b2 ###
%b2 #&#
\bibitem[\protect\citeauthoryear{Almasy et~al.}{2011}]{Almasyetal}
%
\begin{barticle}[pbm]
\bauthor{\bsnm{Almasy},~\bfnm{Laura}\binits{L.}},
\bauthor{\bsnm{Dyer},~\bfnm{Thomas~D.}\binits{T.~D.}},
\bauthor{\bsnm{Peralta},~\bfnm{Juan~Manuel}\binits{J.~M.}},
\bauthor{\bsnm{Kent},~\bfnm{Jack W.}\binits{J.~W.}},
\bauthor{\bsnm{Charlesworth},~\bfnm{Jac~C.}\binits{J.~C.}},
\bauthor{\bsnm{Curran},~\bfnm{Joanne~E.}\binits{J.~E.}} \AND
\bauthor{\bsnm{Blangero},~\bfnm{John}\binits{J.}}
(\byear{2011}).
\btitle{Genetic Analysis Workshop 17 mini-exome simulation}.
\bjournal{BMC Proc.}
\bvolume{5 Suppl 9}
\bpages{S2}.
\bid{doi={10.1186/1753-6561-5-S9-S2}, issn={1753-6561},
pii={1753-6561-5-S9-S2}, pmcid={3287854}, pmid={22373155}}
\end{barticle}
%
\bptok{imsref}%
% NOT OUTPUTED:
% issn = 1753-6561
% fjournal = BMC proceedings
\endbibitem

%b3 ###
%b3 #&#
\bibitem[\protect\citeauthoryear{Asimit and Zeggini}{2010}]{AsimitandZeggini}
%
\begin{barticle}[pbm]
\bauthor{\bsnm{Asimit},~\bfnm{Jennifer}\binits{J.}} \AND
\bauthor{\bsnm{Zeggini},~\bfnm{Eleftheria}\binits{E.}}
(\byear{2010}).
\btitle{Rare variant association analysis methods for complex traits}.
\bjournal{Annu. Rev. Genet.}
\bvolume{44}
\bpages{293--308}.
\bid{doi={10.1146/annurev-genet-102209-163421}, issn={1545-2948},
pmid={21047260}}
\end{barticle}
%
\bptok{imsref}%
% NOT OUTPUTED:
% issn = 1545-2948
% fjournal = Annual review of genetics
\endbibitem

%b4 ###
%b4 #&#
\bibitem[\protect\citeauthoryear{Bansal et~al.}{2010}]{Bansaletal}
%
\begin{barticle}[pbm]
\bauthor{\bsnm{Bansal},~\bfnm{Vikas}\binits{V.}},
\bauthor{\bsnm{Libiger},~\bfnm{Ondrej}\binits{O.}},
\bauthor{\bsnm{Torkamani},~\bfnm{Ali}\binits{A.}} \AND
\bauthor{\bsnm{Schork},~\bfnm{Nicholas~J.}\binits{N.~J.}}
(\byear{2010}).
\btitle{Statistical analysis strategies for association studies
involving rare variants}.
\bjournal{Nat. Rev. Genet.}
\bvolume{11}
\bpages{773--785}.
\bid{doi={10.1038/nrg2867}, issn={1471-0064}, mid={NIHMS497417},
pii={nrg2867}, pmcid={3743540}, pmid={20940738}}
\end{barticle}
%
\bptok{imsref}%
% NOT OUTPUTED:
% issn = 1471-0064
% number = 11
% fjournal = Nature reviews. Genetics
\endbibitem

%b5 ###
%b5 #&#
\bibitem[\protect\citeauthoryear{Barnett, Lee and Lin}{2013}]{Burnett}
%
\begin{barticle}[author]
\bauthor{\bsnm{Barnett},~\bfnm{Ian~J.}\binits{I.~J.}},
\bauthor{\bsnm{Lee},~\bfnm{Seunggeun}\binits{S.}} \AND
\bauthor{\bsnm{Lin},~\bfnm{Xihong}\binits{X.}}
(\byear{2013}).
\btitle{Detecting rare variant effects using extreme phenotype sampling
in sequencing association studies}.
\bjournal{Genet. Epidemiol.}
\bvolume{37}
\bpages{142--151}.
\end{barticle}
%
\bptok{imsref}%
\endbibitem

%b6 ###
%b6 #&#
\bibitem[\protect\citeauthoryear{Basu and Pan}{2011}]{BasuandPan}
%
\begin{barticle}[pbm]
\bauthor{\bsnm{Basu},~\bfnm{Saonli}\binits{S.}} \AND
\bauthor{\bsnm{Pan},~\bfnm{Wei}\binits{W.}}
(\byear{2011}).
\btitle{Comparison of statistical tests for disease association with
rare variants}.
\bjournal{Genet. Epidemiol.}
\bvolume{35}
\bpages{606--619}.
\bid{doi={10.1002/gepi.20609}, issn={1098-2272}, mid={NIHMS308067},
pmcid={3197766}, pmid={21769936}}
\end{barticle}
%
\bptok{imsref}%
% NOT OUTPUTED:
% issn = 1098-2272
% number = 7
% fjournal = Genetic epidemiology
\endbibitem

%b7 ###
%b7 #&#
\bibitem[\protect\citeauthoryear{Daye, Li and Wei}{2012}]{Dayeetal.}
%
\begin{barticle}[pbm]
\bauthor{\bsnm{Daye},~\bfnm{Z.~John}\binits{Z.~J.}},
\bauthor{\bsnm{Li},~\bfnm{Hongzhe}\binits{H.}} \AND
\bauthor{\bsnm{Wei},~\bfnm{Zhi}\binits{Z.}}
(\byear{2012}).
\btitle{A powerful test for multiple rare variants association studies
that incorporates sequencing qualities}.
\bjournal{Nucleic Acids Res.}
\bvolume{40}
\bpages{e60}.
\bid{doi={10.1093/nar/gks024}, issn={1362-4962}, pii={gks024},
pmcid={3340416}, pmid={22262732}}
\end{barticle}
%
\bptok{imsref}%
% NOT OUTPUTED:
% issn = 1362-4962
% number = 8
% fjournal = Nucleic acids research
\endbibitem

%b8 ###
%b8 #&#
\bibitem[\protect\citeauthoryear{Derkach, Lawless and Sun}{2013a}]{Supp}
%
\begin{bmisc}[author]
\bauthor{\bsnm{Derkach},~\bfnm{A.}\binits{A.}},
\bauthor{\bsnm{Lawless},~\bfnm{J~F.}\binits{J.~F.}} \AND
\bauthor{\bsnm{Sun},~\bfnm{L.}\binits{L.}}
(\byear{2013}a).
\bhowpublished{Supplement to ``Pooled association tests for rare genetic
variants: A review and some new results.''
DOI:\doiurl{10.1214/13-STS456SUPP}.}
\end{bmisc}
%
\bptok{imsref}%
\endbibitem

%b9 ###
%b9 #&#
\bibitem[\protect\citeauthoryear{Derkach, Lawless and Sun}{2013b}]{derkach}
%
\begin{barticle}[pbm]
\bauthor{\bsnm{Derkach},~\bfnm{Andriy}\binits{A.}},
\bauthor{\bsnm{Lawless},~\bfnm{Jerry~F.}\binits{J.~F.}} \AND
\bauthor{\bsnm{Sun},~\bfnm{Lei}\binits{L.}}
(\byear{2013}b).
\btitle{Robust and powerful tests for rare variants using Fisher's
method to combine evidence of association from two or more
complementary tests}.
\bjournal{Genet. Epidemiol.}
\bvolume{37}
\bpages{110--121}.
\bid{doi={10.1002/gepi.21689}, issn={1098-2272}, pmid={23032573}}
\end{barticle}
%
\bptok{imsref}%
% NOT OUTPUTED:
% issn = 1098-2272
% number = 1
% fjournal = Genetic epidemiology
\endbibitem

%b10 ###
%b10 #&#
\bibitem[\protect\citeauthoryear{Derkach et~al.}{2014}]{gaw18}
%
\begin{bmisc}[author]
\bauthor{\bsnm{Derkach},~\bfnm{A.}\binits{A.}},
\bauthor{\bsnm{Lawless},~\bfnm{J~F.}\binits{J.~F.}},
\bauthor{\bsnm{Merico},~\bfnm{D.}\binits{D.}},
\bauthor{\bsnm{Paterson},~\bfnm{A.~D.}\binits{A.~D.}} \AND
\bauthor{\bsnm{Sun},~\bfnm{L.}\binits{L.}}
(\byear{2014}).
\bhowpublished{Evaluation of gene-based association tests for analyzing rare
variants using Genetic Analysis Workshop 18 data.
\textit{BMC Proc.} \textbf{8 Suppl 1} S9}.
\end{bmisc}
%
\bptok{imsref}%
\endbibitem

%b11 ###
%b11 #&#
\bibitem[\protect\citeauthoryear{Duchesne and Lafaye~de
Micheaux}{2010}]{DuchesneanddeMicheaux}
%
\begin{barticle}[mr]
\bauthor{\bsnm{Duchesne},~\bfnm{Pierre}\binits{P.}} \AND
\bauthor{\bsnm{Lafaye~de Micheaux},~\bfnm{Pierre}\binits{P.}}
(\byear{2010}).
\btitle{Computing the distribution of quadratic forms: Further
comparisons between the {L}iu--{T}ang--{Z}hang approximation and exact methods}.
\bjournal{Comput. Statist. Data Anal.}
\bvolume{54}
\bpages{858--862}.
\bid{doi={10.1016/j.csda.2009.11.025}, issn={0167-9473}, mr={2580921}}
\end{barticle}
%
\bptok{imsref}%
% NOT OUTPUTED:
% issn = 0167-9473
% url = http://dx.doi.org/10.1016/j.csda.2009.11.025
% number = 4
% fjournal = Computational Statistics \& Data Analysis
\endbibitem

%b12 ###
%b12 #&#
\bibitem[\protect\citeauthoryear{Goeman, van~de Geer and van
Houwelingen}{2006}]{Goemanetal}
%
\begin{barticle}[mr]
\bauthor{\bsnm{Goeman},~\bfnm{Jelle~J.}\binits{J.~J.}},
\bauthor{\bparticle{van~de}~\bsnm{Geer},~\bfnm{Sara~A.}\binits{S.~A.}} \AND
\bauthor{\bparticle{van}~\bsnm{Houwelingen},~\bfnm{Hans~C.}\binits{H.~C.}}
(\byear{2006}).
\btitle{Testing against a high dimensional alternative}.
\bjournal{J. R. Stat. Soc. Ser. B Stat. Methodol.}
\bvolume{68}
\bpages{477--493}.
\bid{doi={10.1111/j.1467-9868.2006.00551.x}, issn={1369-7412}, mr={2278336}}
\end{barticle}
%
\bptok{imsref}%
% NOT OUTPUTED:
% issn = 1369-7412
% url = http://dx.doi.org/10.1111/j.1467-9868.2006.00551.x
% number = 3
% fjournal = Journal of the Royal Statistical Society. Series B.
%Statistical Methodology
\endbibitem

%b13 ###
%b13 #&#
\bibitem[\protect\citeauthoryear{Han and Pan}{2010}]{HanandPan}
%
\begin{barticle}[pbm]
\bauthor{\bsnm{Han},~\bfnm{Fang}\binits{F.}} \AND
\bauthor{\bsnm{Pan},~\bfnm{Wei}\binits{W.}}
(\byear{2010}).
\btitle{A data-adaptive sum test for disease association with multiple
common or rare variants}.
\bjournal{Hum. Hered.}
\bvolume{70}
\bpages{42--54}.
\bid{doi={10.1159/000288704}, issn={1423-0062}, pii={000288704},
pmcid={2912645}, pmid={20413981}}
\end{barticle}
%
\bptok{imsref}%
% NOT OUTPUTED:
% issn = 1423-0062
% number = 1
% fjournal = Human heredity
\endbibitem

%b14 ###
%b14 #&#
\bibitem[\protect\citeauthoryear{Hindorff et~al.}{2009}]{Hindorffetal}
%
\begin{barticle}[author]
\bauthor{\bsnm{Hindorff},~\bfnm{Lucia~A.}\binits{L.~A.}},
\bauthor{\bsnm{Sethupathy},~\bfnm{Praveen}\binits{P.}},
\bauthor{\bsnm{Junkins},~\bfnm{Heather~A.}\binits{H.~A.}},
\bauthor{\bsnm{Ramos},~\bfnm{Erin~M.}\binits{E.~M.}},
\bauthor{\bsnm{Mehta},~\bfnm{Jayashri~P.}\binits{J.~P.}},
\bauthor{\bsnm{Collins},~\bfnm{Francis~S.}\binits{F.~S.}} \AND
\bauthor{\bsnm{Manolio},~\bfnm{Teri~A.}\binits{T.~A.}}
(\byear{2009}).
\btitle{{Potential etiologic and functional implications of genome-wide
association loci for human diseases and traits}}.
\bjournal{Proc. Natl. Acad. Sci. USA}
\bvolume{106}
\bpages{9362--9367}.
\end{barticle}
%
\bptok{imsref}%
\endbibitem

%b15 ###
%b15 #&#
\bibitem[\protect\citeauthoryear{Hoffmann, Marini and
Witte}{2010}]{Hoffmannetal}
%
\begin{barticle}[author]
\bauthor{\bsnm{Hoffmann},~\bfnm{Thomas~J.}\binits{T.~J.}},
\bauthor{\bsnm{Marini},~\bfnm{Nicholas~J.}\binits{N.~J.}} \AND
\bauthor{\bsnm{Witte},~\bfnm{John~S.}\binits{J.~S.}}
(\byear{2010}).
\btitle{Comprehensive approach to analyzing rare genetic variants}.
\bjournal{PLoS ONE}
\bvolume{5}
\bpages{e13584}.
\end{barticle}
%
\bptok{imsref}%
\endbibitem

%b16 ###
%b16 #&#
\bibitem[\protect\citeauthoryear{Huang and Lin}{2007}]{Huangandlin}
%
\begin{barticle}[pbm]
\bauthor{\bsnm{Huang},~\bfnm{B.~E.}\binits{B.~E.}} \AND
\bauthor{\bsnm{Lin},~\bfnm{D.~Y.}\binits{D.~Y.}}
(\byear{2007}).
\btitle{Efficient association mapping of quantitative trait loci with
selective genotyping}.
\bjournal{Am. J. Hum. Genet.}
\bvolume{80}
\bpages{567--576}.
\bid{doi={10.1086/512727}, issn={0002-9297},
pii={S0002-9297(07)60107-4}, pmcid={1821103}, pmid={17273979}}
\end{barticle}
%
\bptok{imsref}%
% NOT OUTPUTED:
% issn = 0002-9297
% number = 3
% fjournal = American journal of human genetics
\endbibitem

%b17 ###
%b17 #&#
\bibitem[\protect\citeauthoryear{King, Rathouz and Nicolae}{2010}]{Kingetal}
%
\begin{barticle}[pbm]
\bauthor{\bsnm{King},~\bfnm{C.~Ryan}\binits{C.~R.}},
\bauthor{\bsnm{Rathouz},~\bfnm{Paul~J.}\binits{P.~J.}} \AND
\bauthor{\bsnm{Nicolae},~\bfnm{Dan~L.}\binits{D.~L.}}
(\byear{2010}).
\btitle{An evolutionary framework for association testing in
resequencing studies}.
\bjournal{PLoS Genet.}
\bvolume{6}
\bpages{e1001202}.
\bid{doi={10.1371/journal.pgen.1001202}, issn={1553-7404},
pmcid={2978703}, pmid={21085648}}
\end{barticle}
%
\bptok{imsref}%
% NOT OUTPUTED:
% issn = 1553-7404
% number = 11
% fjournal = PLoS genetics
\endbibitem

%b18 ###
%b18 #&#
\bibitem[\protect\citeauthoryear{Ladouceur et~al.}{2012}]{Ladouceur}
%
\begin{barticle}[pbm]
\bauthor{\bsnm{Ladouceur},~\bfnm{Martin}\binits{M.}},
\bauthor{\bsnm{Dastani},~\bfnm{Zari}\binits{Z.}},
\bauthor{\bsnm{Aulchenko},~\bfnm{Yurii~S.}\binits{Y.~S.}},
\bauthor{\bsnm{Greenwood},~\bfnm{Celia~M.~T.}\binits{C.~M.~T.}} \AND
\bauthor{\bsnm{Richards},~\bfnm{J.~Brent}\binits{J.~B.}}
(\byear{2012}).
\btitle{The empirical power of rare variant association methods:
Results from sanger sequencing in 1998 individuals}.
\bjournal{PLoS Genet.}
\bvolume{8}
\bpages{e1002496}.
\bid{doi={10.1371/journal.pgen.1002496}, issn={1553-7404},
pii={PGENETICS-D-11-01087}, pmcid={3271058}, pmid={22319458}}
\end{barticle}
%
\bptok{imsref}%
% NOT OUTPUTED:
% issn = 1553-7404
% number = 2
% fjournal = PLoS genetics
\endbibitem

%b19 ###
%b19 #&#
\bibitem[\protect\citeauthoryear{Lee, Wu and Lin}{2012}]{Leeetal}
%
\begin{barticle}[pbm]
\bauthor{\bsnm{Lee},~\bfnm{Seunggeun}\binits{S.}},
\bauthor{\bsnm{Wu},~\bfnm{Michael~C.}\binits{M.~C.}} \AND
\bauthor{\bsnm{Lin},~\bfnm{Xihong}\binits{X.}}
(\byear{2012}).
\btitle{Optimal tests for rare variant effects in sequencing
association studies}.
\bjournal{Biostatistics}
\bvolume{13}
\bpages{762--775}.
\bid{doi={10.1093/biostatistics/kxs014}, issn={1468-4357},
pii={kxs014}, pmcid={3440237}, pmid={22699862}}
\end{barticle}
%
\bptok{imsref}%
% NOT OUTPUTED:
% issn = 1468-4357
% number = 4
% fjournal = Biostatistics (Oxford, England)
\endbibitem

%b20 ###
%b20 #&#
\bibitem[\protect\citeauthoryear{Li and Lagakos}{2006}]{LiandLagakos}
%
\begin{barticle}[mr]
\bauthor{\bsnm{Li},~\bfnm{Qian~H.}\binits{Q.~H.}} \AND
\bauthor{\bsnm{Lagakos},~\bfnm{Stephen~W.}\binits{S.~W.}}
(\byear{2006}).
\btitle{On the relationship between directional and omnibus statistical tests}.
\bjournal{Scand. J. Stat.}
\bvolume{33}
\bpages{239--246}.
\bid{doi={10.1111/j.1467-9469.2005.00489.x}, issn={0303-6898}, mr={2279640}}
\end{barticle}
%
\bptok{imsref}%
% NOT OUTPUTED:
% issn = 0303-6898
% url = http://dx.doi.org/10.1111/j.1467-9469.2005.00489.x
% number = 2
% fjournal = Scandinavian Journal of Statistics. Theory and Applications
\endbibitem

%b21 ###
%b21 #&#
\bibitem[\protect\citeauthoryear{Li and Leal}{2008}]{LiandLeal}
%
\begin{barticle}[pbm]
\bauthor{\bsnm{Li},~\bfnm{Bingshan}\binits{B.}} \AND
\bauthor{\bsnm{Leal},~\bfnm{Suzanne~M.}\binits{S.~M.}}
(\byear{2008}).
\btitle{Methods for detecting associations with rare variants for
common diseases: Application to analysis of sequence data}.
\bjournal{Am. J. Hum. Genet.}
\bvolume{83}
\bpages{311--321}.
\bid{doi={10.1016/j.ajhg.2008.06.024}, issn={1537-6605},
pii={S0002-9297(08)00408-4}, pmcid={2842185}, pmid={18691683}}
\end{barticle}
%
\bptok{imsref}%
% NOT OUTPUTED:
% issn = 1537-6605
% number = 3
% fjournal = American journal of human genetics
\endbibitem

%b22 ###
%b22 #&#
\bibitem[\protect\citeauthoryear{Lin and Tang}{2011}]{LinandTang}
%
\begin{barticle}[author]
\bauthor{\bsnm{Lin},~\bfnm{Dan-Yu}\binits{D.-Y.}} \AND
\bauthor{\bsnm{Tang},~\bfnm{Zheng-Zheng}\binits{Z.-Z.}}
(\byear{2011}).
\btitle{A general framework for detecting disease associations with
rare variants in sequencing studies}.
\bjournal{The American Journal of Human Genetics}
\bvolume{89}
\bpages{354--367}.
\end{barticle}
%
\bptok{imsref}%
\endbibitem

%b23 ###
%b23 #&#
\bibitem[\protect\citeauthoryear{Madsen and Browning}{2009}]{MadsenandBrowning}
%
\begin{barticle}[pbm]
\bauthor{\bsnm{Madsen},~\bfnm{Bo~Eskerod}\binits{B.~E.}} \AND
\bauthor{\bsnm{Browning},~\bfnm{Sharon~R.}\binits{S.~R.}}
(\byear{2009}).
\btitle{A groupwise association test for rare mutations using a
weighted sum statistic}.
\bjournal{PLoS Genet.}
\bvolume{5}
\bpages{e1000384}.
\bid{doi={10.1371/journal.pgen.1000384}, issn={1553-7404},
pmcid={2633048}, pmid={19214210}}
\end{barticle}
%
\bptok{imsref}%
% NOT OUTPUTED:
% issn = 1553-7404
% number = 2
% fjournal = PLoS genetics
\endbibitem

%b24 ###
%b24 #&#
\bibitem[\protect\citeauthoryear{Manolio, Brooks and
Collins}{2008}]{Manolioetal2008}
%
\begin{barticle}[pbm]
\bauthor{\bsnm{Manolio},~\bfnm{Teri~A.}\binits{T.~A.}},
\bauthor{\bsnm{Brooks},~\bfnm{Lisa~D.}\binits{L.~D.}} \AND
\bauthor{\bsnm{Collins},~\bfnm{Francis~S.}\binits{F.~S.}}
(\byear{2008}).
\btitle{A~HapMap harvest of insights into the genetics of common disease}.
\bjournal{J. Clin. Invest.}
\bvolume{118}
\bpages{1590--1605}.
\bid{doi={10.1172/JCI34772}, issn={0021-9738}, pmcid={2336881}, pmid={18451988}}
\end{barticle}
%
\bptok{imsref}%
% NOT OUTPUTED:
% issn = 0021-9738
% number = 5
% fjournal = The Journal of clinical investigation
\endbibitem

%b25 ###
%b25 #&#
\bibitem[\protect\citeauthoryear{Mardia, Kent and Bibby}{1979}]{Mardia}
%
\begin{bbook}[author]
\bauthor{\bsnm{Mardia},~\bfnm{K.~V.}\binits{K.~V.}},
\bauthor{\bsnm{Kent},~\bfnm{J.~T.}\binits{J.~T.}} \AND
\bauthor{\bsnm{Bibby},~\bfnm{J.~M.}\binits{J.~M.}}
(\byear{1979}).
\btitle{Multivariate Analysis}.
\bpublisher{Academic Press},
\blocation{Waltham, MA}.
\end{bbook}
%
\bptok{imsref}%
\endbibitem

%b26 ###
%b26 #&#
\bibitem[\protect\citeauthoryear{Morgenthaler and
Thilly}{2007}]{MorgenthalerandThilly}
%
\begin{barticle}[author]
\bauthor{\bsnm{Morgenthaler},~\bfnm{Stephan}\binits{S.}} \AND
\bauthor{\bsnm{Thilly},~\bfnm{William~G.}\binits{W.~G.}}
(\byear{2007}).
\btitle{A strategy to discover genes that carry multi-allelic or
mono-allelic risk for common diseases: A cohort allelic sums test (CAST)}.
\bjournal{Mutation Research/Fundamental and Molecular Mechanisms of Mutagenesis}
\bvolume{615}
\bpages{28--56}.
\end{barticle}
%
\bptok{imsref}%
\endbibitem

%b27 ###
%b27 #&#
\bibitem[\protect\citeauthoryear{Morris and Zeggini}{2010}]{MorrisandZeggini}
%
\begin{barticle}[pbm]
\bauthor{\bsnm{Morris},~\bfnm{Andrew~P.}\binits{A.~P.}} \AND
\bauthor{\bsnm{Zeggini},~\bfnm{Eleftheria}\binits{E.}}
(\byear{2010}).
\btitle{An evaluation of statistical approaches to rare variant
analysis in genetic association studies}.
\bjournal{Genet. Epidemiol.}
\bvolume{34}
\bpages{188--193}.
\bid{doi={10.1002/gepi.20450}, issn={1098-2272}, pmcid={2962811},
pmid={19810025}}
\end{barticle}
%
\bptok{imsref}%
% NOT OUTPUTED:
% issn = 1098-2272
% number = 2
% fjournal = Genetic epidemiology
\endbibitem

%b28 ###
%b28 #&#
\bibitem[\protect\citeauthoryear{Neale et~al.}{2011}]{Nealeetal}
%
\begin{barticle}[author]
\bauthor{\bsnm{Neale},~\bfnm{Benjamin~M.}\binits{B.~M.}},
\bauthor{\bsnm{Rivas},~\bfnm{Manuel~A.}\binits{M.~A.}},
\bauthor{\bsnm{Voight},~\bfnm{Benjamin~F.}\binits{B.~F.}},
\bauthor{\bsnm{Altshuler},~\bfnm{David}\binits{D.}} \betal{et al.}
(\byear{2011}).
\btitle{Testing for an unusual distribution of rare variants}.
\bjournal{PLoS Genet.}
\bvolume{7}
\bpages{e1001322}.
\end{barticle}
%
\bptok{imsref}%
\endbibitem

%b29 ###
%b29 #&#
\bibitem[\protect\citeauthoryear{Owen}{2009}]{Owen}
%
\begin{barticle}[mr]
\bauthor{\bsnm{Owen},~\bfnm{Art~B.}\binits{A.~B.}}
(\byear{2009}).
\btitle{Karl {P}earson's meta-analysis revisited}.
\bjournal{Ann. Statist.}
\bvolume{37}
\bpages{3867--3892}.
\bid{doi={10.1214/09-AOS697}, issn={0090-5364}, mr={2572446}}
\end{barticle}
%
\bptok{imsref}%
% NOT OUTPUTED:
% issn = 0090-5364
% url = http://dx.doi.org/10.1214/09-AOS697
% number = 6B
% coden = ASTSC7
% fjournal = The Annals of Statistics
\endbibitem\vadjust{\goodbreak}

%b30 ###
%b30 #&#
\bibitem[\protect\citeauthoryear{Pan}{2009}]{Pan}
%
\begin{barticle}[pbm]
\bauthor{\bsnm{Pan},~\bfnm{Wei}\binits{W.}}
(\byear{2009}).
\btitle{Asymptotic tests of association with multiple SNPs in linkage
disequilibrium}.
\bjournal{Genet. Epidemiol.}
\bvolume{33}
\bpages{497--507}.
\bid{doi={10.1002/gepi.20402}, issn={1098-2272}, mid={NIHMS93959},
pmcid={2732754}, pmid={19170135}}
\end{barticle}
%
\bptok{imsref}%
% NOT OUTPUTED:
% issn = 1098-2272
% number = 6
% fjournal = Genetic epidemiology
\endbibitem

%b31 ###
%b31 #&#
\bibitem[\protect\citeauthoryear{Price et~al.}{2010}]{Priceetal}
%
\begin{barticle}[author]
\bauthor{\bsnm{Price},~\bfnm{Alkes~L.}\binits{A.~L.}},
\bauthor{\bsnm{Kryukov},~\bfnm{Gregory~V.}\binits{G.~V.}},
\bauthor{\bparticle{de} \bsnm{Bakker},~\bfnm{Paul~I.}\binits{P.~I.}},
\bauthor{\bsnm{Purcell},~\bfnm{Shaun~M.}\binits{S.~M.}} \betal{et al.}
(\byear{2010}).
\btitle{{Pooled association tests for rare variants in
exon-resequencing studies.}}
\bjournal{The American Journal of Human Genetics}
\bvolume{86}
\bpages{832--838}.
\end{barticle}
%
\bptok{imsref}%
\endbibitem

%b32 ###
%b32 #&#
\bibitem[\protect\citeauthoryear{Rao}{1973}]{Rao}
%
\begin{bbook}[mr]
\bauthor{\bsnm{Rao},~\bfnm{C.~Radhakrishna}\binits{C.~R.}}
(\byear{1973}).
\btitle{Linear Statistical Inference and Its Applications},
\bedition{2nd} ed.
\bpublisher{Wiley},
\blocation{Hoboken, NJ}.
\bid{mr={0346957}}
\end{bbook}
%
\bptok{imsref}%
% NOT OUTPUTED:
% fpage = xx+625
\endbibitem

%b33 ###
%b33 #&#
\bibitem[\protect\citeauthoryear{Reich et~al.}{2001}]{Reich}
%
\begin{barticle}[pbm]
\bauthor{\bsnm{Reich},~\bfnm{D.~E.}\binits{D.~E.}},
\bauthor{\bsnm{Cargill},~\bfnm{M.}\binits{M.}},
\bauthor{\bsnm{Bolk},~\bfnm{S.}\binits{S.}},
\bauthor{\bsnm{Ireland},~\bfnm{J.}\binits{J.}},
\bauthor{\bsnm{Sabeti},~\bfnm{P.~C.}\binits{P.~C.}},
\bauthor{\bsnm{Richter},~\bfnm{D.~J.}\binits{D.~J.}},
\bauthor{\bsnm{Lavery},~\bfnm{T.}\binits{T.}},
\bauthor{\bsnm{Kouyoumjian},~\bfnm{R.}\binits{R.}},
\bauthor{\bsnm{Farhadian},~\bfnm{S.~F.}\binits{S.~F.}},
\bauthor{\bsnm{Ward},~\bfnm{R.}\binits{R.}} \AND
\bauthor{\bsnm{Lander},~\bfnm{E.~S.}\binits{E.~S.}}
(\byear{2001}).
\btitle{Linkage disequilibrium in the human genome}.
\bjournal{Nature}
\bvolume{411}
\bpages{199--204}.
\bid{doi={10.1038/35075590}, issn={0028-0836}, pii={35075590}, pmid={11346797}}
\end{barticle}
%
\bptok{imsref}%
% NOT OUTPUTED:
% issn = 0028-0836
% number = 6834
\endbibitem

%b34 ###
%b34 #&#
\bibitem[\protect\citeauthoryear{Skotte, Korneliussen and
Albrechtsen}{2012}]{Skotteetal.}
%
\begin{barticle}[pbm]
\bauthor{\bsnm{Skotte},~\bfnm{Line}\binits{L.}},
\bauthor{\bsnm{Korneliussen},~\bfnm{Thorfinn~Sand}\binits{T.~S.}} \AND
\bauthor{\bsnm{Albrechtsen},~\bfnm{Anders}\binits{A.}}
(\byear{2012}).
\btitle{Association testing for next-generation sequencing data using
score statistics}.
\bjournal{Genet. Epidemiol.}
\bvolume{36}
\bpages{430--437}.
\bid{doi={10.1002/gepi.21636}, issn={1098-2272}, pmid={22570057}}
\end{barticle}
%
\bptok{imsref}%
% NOT OUTPUTED:
% issn = 1098-2272
% number = 5
% fjournal = Genetic epidemiology
\endbibitem

\bibitem[\protect\citeauthoryear{Sul, Buhm  and Eleazar}{2011}]{Su}
%
\begin{barticle}[author]
\bauthor{\bsnm{Sul},~\bfnm{J. H.}\binits{J. H.}},
\bauthor{\bsnm{Buhm},~\bfnm{H.}\binits{H.}} \AND
\bauthor{\bsnm{Eleazar},~\bfnm{E.}\binits{E.}}
(\byear{2011}).
\btitle{Increasing power of groupwise association
test with likelihood ratio test}.
\bjournal{J. Comput. Biol.}
\bvolume{18}
\bpages{1611--1624}.
\end{barticle}
%
\bptok{imsref}%
\endbibitem


%b35 ###
%b35 #&#
\bibitem[\protect\citeauthoryear{Wu et~al.}{2011}]{Wuetal}
%
\begin{barticle}[author]
\bauthor{\bsnm{Wu},~\bfnm{Michael~C.}\binits{M.~C.}},
\bauthor{\bsnm{Lee},~\bfnm{Seunggeun}\binits{S.}},
\bauthor{\bsnm{Cai},~\bfnm{Tianxi}\binits{T.}},
\bauthor{\bsnm{Li},~\bfnm{Yun}\binits{Y.}},
\bauthor{\bsnm{Boehnke},~\bfnm{Michael}\binits{M.}} \AND
\bauthor{\bsnm{Lin},~\bfnm{Xihong}\binits{X.}}
(\byear{2011}).
\btitle{{Rare-variant association testing for sequencing data with the
sequence Kernel association test}}.
\bjournal{The American Journal of Human Genetics}
\bvolume{89}
\bpages{82--93}.
\end{barticle}
%
\bptok{imsref}%
\endbibitem

%b36 ###
%b36 #&#
\bibitem[\protect\citeauthoryear{Yi and Zhi}{2011}]{YiandZhi}
%
\begin{barticle}[pbm]
\bauthor{\bsnm{Yi},~\bfnm{Nengjun}\binits{N.}} \AND
\bauthor{\bsnm{Zhi},~\bfnm{Degui}\binits{D.}}
(\byear{2011}).
\btitle{Bayesian analysis of rare variants in genetic association studies}.
\bjournal{Genet. Epidemiol.}
\bvolume{35}
\bpages{57--69}.
\bid{doi={10.1002/gepi.20554}, issn={1098-2272}, mid={NIHMS331121},
pmcid={3200544}, pmid={21181897}}
\end{barticle}
%
\bptok{imsref}%
% NOT OUTPUTED:
% issn = 1098-2272
% number = 1
% fjournal = Genetic epidemiology
\endbibitem

%b37 ###
%b37 #&#
\bibitem[\protect\citeauthoryear{Yilmaz and Bull}{2011}]{YilmazandBull}
%
\begin{barticle}[pbm]
\bauthor{\bsnm{Yilmaz},~\bfnm{Yildiz~E.}\binits{Y.~E.}} \AND
\bauthor{\bsnm{Bull},~\bfnm{Shelley~B.}\binits{S.~B.}}
(\byear{2011}).
\btitle{Are quantitative trait-dependent sampling designs
cost-effective for analysis of rare and common variants?}
\bjournal{BMC Proc.}
\bvolume{5 Suppl 9}
\bpages{S111}.
\bid{doi={10.1186/1753-6561-5-S9-S111}, issn={1753-6561},
pii={1753-6561-5-S9-S111}, pmcid={3287835}, pmid={22373146}}
\end{barticle}
%
\bptok{imsref}%
% NOT OUTPUTED:
% issn = 1753-6561
% fjournal = BMC proceedings
\endbibitem

\end{thebibliography}
\end{document}